\documentclass[submission, Phys]{SciPost}
\usepackage[T1]{fontenc}
\usepackage{booktabs}
\usepackage{graphicx}
\usepackage{xspace}
\usepackage{multirow}
\usepackage{amssymb,amsmath}
\usepackage{mathtools}
\usepackage{cleveref}
\usepackage{xstring}
\usepackage[normalem]{ulem}
\usepackage{algorithm}
\usepackage{algpseudocode}
\usepackage{ulem}
\usepackage{mathrsfs}
\usepackage{calligra}
\hypersetup{colorlinks, linkcolor={red!50!black}, citecolor={blue!50!black},
  urlcolor={blue!80!black}}

\newcommand{\trash}[1]{}

\begin{document}

\thispagestyle{empty}

\begin{center}

\begin{Large}
\textbf{\textsc{Di-Higgs to 4$b$ with Bayesian inference:\\ improving simulation estimates}}

\end{Large}

\vspace{1cm}

{\sc
Ezequiel~Alvarez$^{a}$%
\footnote{{\tt \href{mailto:sequi@unsam.edu.ar}{sequi@unsam.edu.ar}}}%
, Leandro~Da~Rold$^{b}$%
\footnote{{\tt \href{mailto:daroldl@ib.edu.ar}{daroldl@ib.edu.ar}}}%
, Manuel~Szewc$^{a}$%
\footnote{{\tt \href{mailto:szewcml@ucmail.uc.edu}{mszewc@unsam.edu.ar}}}%
, Alejandro~Szynkman$^{c}$%
\footnote{{\tt \href{mailto:szynkman@fisica.unlp.edu.ar}{szynkman@fisica.unlp.edu.ar}}}%
, Santiago~Tanco$^{c}$%
\footnote{{\tt \href{mailto:santiago.tanco@fisica.unlp.edu.ar}{santiago.tanco@fisica.unlp.edu.ar}}}%
, Tatiana~Tarutina$^{c}$%
\footnote{{\tt \href{mailto:tarutina@fisica.unlp.edu.ar}{tarutina@fisica.unlp.edu.ar}}}%
}

\vspace*{.7cm}

{\sl
$^a$International Center for Advanced Studies (ICAS) and ICIFI-CONICET, UNSAM, \\
25 de Mayo y Francia, CP1650, San Mart\'{\i}n, Buenos Aires, Argentina

\vspace*{0.1cm}
$^b$ Centro Atómico Bariloche, Instituto Balseiro and CONICET, \\
Av. Bustillo 9500, 8400, S.C. de Bariloche, Argentina

\vspace*{0.1cm}

$^c$IFLP, CONICET - Dpto. de F\'{\i}sica, Universidad Nacional de La Plata, \\ 
C.C. 67, 1900 La Plata, Argentina

\vspace*{0.1cm}

}

\end{center}

\vspace{0.1cm}

\begin{abstract}
\noindent 
Measuring di-Higgs production in the four-bottom channel is challenged by overwhelming QCD backgrounds and imperfect simulations. We develop a Bayesian mixture model that simultaneously infers signal and background fractions and their individual shapes directly in the signal region. The likelihood is a nuanced combination of a one-dimensional kinematic discriminator and per-jet flavour scores; with their correlations incorporated via kinematic bins. Monte Carlo informs weak Dirichlet priors, while the posterior adjusts to the interplay of the model, priors and observed data. Using pseudo-data simulated with standard tools and with controlled mismatches, we show that the method corrects biased priors, delivers calibrated 68–95\% credible intervals for the signal count, and improves dataset-level ROC/AUC relative to simple cut-and-count baselines. This study highlights how Bayesian inference can harvest information present in the signal region and self-calibrate model parameters, providing a robust route to increased sensitivity in di-Higgs searches.
\vskip 1cm
~
\end{abstract}

\def\thefootnote{\arabic{footnote}}
\setcounter{page}{0}
\setcounter{footnote}{0}

\newpage

\tableofcontents

\section{Introduction}\label{sec:introduction}

Di-Higgs production is a key target at the LHC because it directly probes the Higgs self-coupling—among the last unmeasured parameters of the Standard Model Higgs sector—and thus the shape of the Higgs potential. Measuring this coupling would provide a stringent test of the mechanism of electroweak symmetry breaking, and any deviation from the Standard Model prediction would be not only a clear sign of New Physics, but also a clue to its underlying dynamics. However, the prospects for measuring this parameter within the LHC program are challenging \cite{LHCHiggsCrossSectionWorkingGroup:2016ypw,DiMicco:2019ngk}. Among the available final states, the so-called `golden channel' is $hh \to b\bar b \gamma\gamma$ which offers a favorable balance between branching fraction and manageable backgrounds, both in rate and systematic uncertainties \cite{ATLAS:2025bsu,CMS:2020tkr}. Alternatively, the $hh \to b\bar bb\bar b$ channel benefits from a roughly two order of magnitude larger branching fraction, but its sensitivity is severely limited by overwhelming QCD backgrounds and their difficult‑to‑control uncertainties~\cite{ATLAS:2018rnh,ATLAS:2023qzf,CMS:2022cpr,CMS:2022gjd,CMS:2025ero}—compounded by the usual practical complications of complex collider analyses.

In this work, and as a part of a larger program \cite{Alvarez:2021hxu,Alvarez:2021zje,Alvarez:2022kfr,Alvarez:2022qoz,Alvarez:2024doi,Alvarez:2024owq,Alvarez:2025cnz}, we aim to exploit Bayesian inference in combining models and data\footnote{Along this article we use data to refer to pseudo-data, since we do not work with real data.} to enhance the capability of the $hh \to b\bar bb\bar b$ channel to measure the Higgs self-coupling.  Bayesian machine learning updates beliefs about models and parameters by combining prior knowledge with new data using Bayes' Theorem. In this article, we aim to show that the substantial uncertainties and potential biases in the $hh \to b\bar b b\bar b$ signal and background, arising from theory, experiment, and Monte Carlo (MC) simulations, can be reduced by developing data-driven models whose parameter distributions are learned directly from data.

Di-Higgs searches are being conducted by the ATLAS and CMS collaborations (see Refs.~\cite{ATLAS:2024ish,CMS:2025ngq} for the respective most recent combined measurements), and various empirical approaches have been proposed to improve sensitivity, both within the Standard Model (SM)~\cite{Chiang:2024pho,Chekanov:2025xpk} and in searches for new physics (NP)~\cite{Huang:2017jws, ATLAS:2024lsk}.
In all cases, the dominant challenge is the overwhelming background with estimation uncertainties that are large compared to the relatively tiny expected signal fraction.  
With recent advances in machine learning, many proposals leverage these techniques to increase sensitivity. 
However, di-Higgs searches present unique challenges. While the most straightforward applications of machine learning to High Energy Physics implement supervised algorithms trained on MC simulations to increase the sensitivity of the analysis, either via improved cuts or by proving an optimal summary statistic~\cite{Plehn:2022ftl}, MC simulations are not able to model the complex multi-jet background relevant to di-Higgs searches and can only provide useful approximations.
Consequently, a neural network trained to very high precision on MC may learn MC-specific artifacts and fail to generalize to real data.  This is a well known issue and there are different workarounds, as for instance in \cite{atlas4b} where the differential background is estimated in signal region by reweighting events from a control region with a neural network reweighter, which is trained in data through additional control regions, effectively implementing a differential ABCD-like method.  Although the method is validated with an independent control region and associated systematic tests, its main limitation is the transferability assumption: the conditional feature distributions—and hence the ratio—must be sufficiently similar between the control and signal regions.   In this manuscript, we adopt a Bayesian machine‑learning approach that proposes to learn directly on signal‑region, where both signal and background are present, and infer their respective contributions directly from data via a probabilistic mixture model defined over the component parameters and mixing fractions.

In previous works, we explored Bayesian methods for mixture models under a conditional independence assumption: the observables are modeled as independent within each class (signal or backgrounds), and any correlation between observables arises from the unobserved marginalization over class assignment. Under this hypothesis, the Bayesian formulation delivered notable sensitivity gains in di‑Higgs searches to $b\bar b\gamma\gamma$~\cite{Alvarez:2022kfr}, in four‑top final states~\cite{4tops}, and in improvements to the traditional ABCD method~\cite{Alvarez:2024owq} for $hh\to b\bar bb\bar b$. In particular, Ref.~\cite{Alvarez:2024doi} demonstrated the effectiveness of conditional independence in a four‑dimensional flavour‑tagging problem for four‑jet events; we build on those results here. Despite their strong performance, these approaches all rely on within‑class independence, an assumption often violated in real data where many observables exhibit non‑negligible dependence. To address this limitation, we have recently implemented tools that extend the Bayesian framework to explicitly handle correlated\footnote{Strictly speaking, “dependent” is the accurate term; for simplicity, and following common community usage, we use “correlation” to refer to statistical dependence throughout this manuscript.} variables~\cite{Alvarez:2025cnz} in boosted top-jets. In the present work, we consider a different approach to incorporate correlated variables to di‑Higgs proposed searches.  Beyond the references discussed above, several recent works have applied Bayesian methods to high energy physics (HEP) analyses (see e.g. \cite{Fowlie:2019ydo,Fowlie:2021zyf,DelDebbio:2021whr,Yallup:2022rjd,Fowlie:2024dgj,Candido:2024hjt,Albert:2024zsh,Costantini:2025wxp}), highlighting a promising direction for future studies. This is signaling that the statistical methodology in HEP is evolving rapidly, with the PDG Statistics Review offering a continually updated overview of ongoing best practices \cite{ParticleDataGroup:2024cfk}.

In the current article we extend the work in Ref.~\cite{Alvarez:2024doi} which shows how to improve the multijet flavour tagging through Bayesian density estimation and apply it to a simplified di-Higgs analysis where we restrict the QCD backgrounds to $b \bar b c \bar c$, $c\bar c c \bar c$ and $b \bar b b \bar b$. This restriction, done for simplicity, ensures we consider the most signal-like backgrounds in flavour space, at the price of neglecting potentially overwhelming backgrounds such as $jjjj$ and $jjb\bar b$ and other subleading backgrounds such as $t\bar t$, single Higgs and EW boson production. We test our framework in simulated events where we can implement a more realistic, if simple, $b$-tagging based on the secondary vertices of the jet, and whose tagging curves we infer from data.  In addition, we also include kinematic observables to be able to distinguish signal di-Higgs from the multijet QCD $b\bar bb\bar b$, which is otherwise identical in flavour information. 

Our model incorporates restricted conditional dependence between kinematics and flavour to more accurately represent the underlying physics. It self-corrects for biased priors on shape parameters when inferring the signal yield.

Analyzing datasets where the di-Higgs represent $0-1.5\%$ of the total number of events, the method infers the signal count with calibrated 68\% credible intervals. It shows improved sensitivity over standard benchmarks in this proof-of-concept study.

This work is organized as follows: in Section~\ref{sec:diHiggs} we provide further details on the simulation and event selection strategy implemented and the definition of the ``data'' and ``simulation'' datasets; in Section~\ref{sec:bayesian_models} we provide a brief introduction to Bayesian mixture models and define the probabilistic structure considered in this work; in Section~\ref{sec:model_comparison} we show how the added flexibility of inferring the shapes of the tagging curves and kinematics distributions allows to correct for MC mismodelling more effectively; in Section~\ref{sec:sensitivity} we show how the proposed strategy outputs meaningful credible intervals and compare the obtained sensitivity with standard techniques to showcase the power of the learned model; we conclude and discuss future prospects in Section~\ref{sec:outlook}.

\section{Double Higgs production at the LHC}\label{sec:diHiggs}

Among the different possible Higgs decay channels, we study the double-Higgs production with both Higgs decaying into a pair of bottom quarks, namely $pp\to hh \to b\bar bb\bar b$. The special relevance of this process lies in being one of the most important to be measured in the forthcoming high luminosity stage of the LHC~\cite{atlas4b, cms4b}. As irreducible background for our signal, we consider the dominant QCD contribution with four bottom quarks ($b \bar b b \bar b$). We also include in the analysis QCD processes with a pair of bottoms and a pair of charms ($b \bar b$$c \bar c$) and four charms ($c \bar c c \bar c$), which at detector level contribute as reducible backgrounds as a result of $c$-jets misidentified as $b$-jets. We do not consider other relevant backgrounds, as other multijet final states, $t\bar t$, single Higgs and EW boson production.

For the double-Higss production we consider the non-resonant dominant process given by gluon fusion at one loop level. We generate signal and background events using {\tt MadGraph5\_aMC@NLO} (MG5)~\cite{Alwall:2011uj,Alwall:2014hca} at a center-of-mass energy of 14 TeV. Higgs decay simulations are performed with {\tt MadSpin}~\cite{Artoisenet:2012st}. We then use {\tt Pythia~8}~\cite{Sjostrand:2006za,Sjostrand:2007gs,Sjostrand:2014zea} for parton showering and hadronization, and {\tt Delphes 3}~\cite{deFavereau:2013fsa} for a fast detector simulation, employing a modified CMS card where the jets are reconstructed with the anti-$k_T$ algorithm~\cite{Cacciari:2008gp} setting $R=0.8$ and demanding $p_{T_j} > 8\;{\rm GeV}$. 

We do not consider the default hard $b$-tagging assignment from {\tt Delphes}, since we want to explicitly model the $b$-tagger probability score and avoid any such hard assignments. To do so, we store all events, with true flavour assigned by {\tt Delphes} by searching for the closest mother parton with $|\eta|<5$ and compute the $b$-tagger probability per jet. Instead of a state-of-the-art tagger, we consider the more easily computable, if suboptimal, number of secondary vertices, $N_{\mathrm{SV}}$.  
The variable $N_{\mathrm{SV}}$ is defined in Ref.~\cite{CMS:2012feb} as the number of tracks within an angular distance $\Delta R \leq 0.3$ of the jet axis, with $p_{T} \geq 1$ GeV, and the transverse impact parameter $2.5 \text{ }\mu\text{m} \leq d_{0} \leq 2.0 \text{ mm}$, where $d_{0}$ is the transverse distance to the primary vertex at the point of closest approach in the transverse plane; and we compute it as in Ref.\cite{Faroughy:2022dyq} by using the {\tt ROOT}~\cite{Brun:1997pa, rene_brun_2019_3895860} output of {\tt Delphes} to study the tracks in each jet.

For the analysis, we apply the following cuts to all the jets (irrespective of their true flavour): $p_T>25$~GeV and $|\eta|<2.5$. We select events with at least 4 jets surviving these cuts. For simplicity, we require that these 4 jets are those originating from the hard process. This cut, inapplicable in a real analysis, simplifies the probabilistic model introduced in Section~\ref{sec:bayesian_models}, but is not strictly necessary and will be relaxed in future work.

The resulting four jets are divided in two pairs, each of which is grouped into a Higgs candidate. The grouping is obtained by minimizing a $\chi^2$ metric,
\begin{equation}
    \chi^2 = (m_1-125\,\mathrm{GeV})^2 + (m_2-125\,\mathrm{GeV})^2 \,,
    \label{eq:chi2}
\end{equation}
where $m_{1,2}$ are the masses of the two Higgs candidates composed of a pair of jets, ordered by the $p_{T}$ of the Higgs candidate. The event is accepted if both masses satisfy $|m_i - 125\,\mathrm{GeV}|<25\,\mathrm{GeV}$. This grouping and selection criteria aims to mimic Ref.~\cite{ATLAS:2023qzf}, albeit without the correction in mass due to detector effects. This selection criterion greatly reduces the acceptance of background simulations and is the main bottleneck in generating large samples.

For all events, we store all four $N_{\mathrm{SV}}$ values, which play the role of the $b$-tagger score, and a list of kinematic features $\mathcal{O}$ that provide further information,
\begin{equation}
    \mathcal{O} = \{ \text{Pairing},m_{1},m_{2}, \Delta\eta_{1},\Delta\eta_{2},\Delta\phi_{1},\Delta\phi_{2},p^{h}_{T,1},p^{h}_{T,2},p^{j}_{T,1},p^{j}_{T,2},p^{j}_{T,3},p^{j}_{T,4}\}\,,
    \label{eq:kine_features}
\end{equation}
where $p^{j}_{T,i}$ is the transverse momentum of each jet, labeled by their transverse momentum ordering. For each pair of jets grouped as a Higgs candidate  $\{m_{i}$, $\Delta\eta_{i}$, $\Delta\phi_{i}$, $p^{h}_{T,i}\}$ are the  invariant mass, difference in $\eta$, difference in $\phi$ and addition of $p_T$ of the jets in pair $i$, respectively. $\text{Pairing}$ denotes which jets are grouped together,
\begin{equation}
    \begin{split}
        \text{Pairing} &= 1 \quad \text{for }h_{1} = \{j_{1},j_{2}\} \text{ and }  h_{2} = \{j_{3},j_{4}\}\,, \\
        &= 2 \quad \text{for }h_{1} = \{j_{1},j_{3}\} \text{ and }  h_{2} = \{j_{2},j_{4}\}\,, \\
        &= 3 \quad \text{for }h_{1} = \{j_{1},j_{4}\} \text{ and }  h_{2} = \{j_{2},j_{3}\} \,,
    \end{split}
\end{equation}
where again the jets are labeled according to their $p_{T}$.

This procedure is repeated for two different {\tt Pythia} settings, which act as ``simulation'' and ``measurement'' or ``data'', respectively. In both cases, we turn off Multi-Parton interaction (MPI) for speed, but vary the parton shower. ``Data'' considers the {\tt VINCIA}~\cite{Brooks:2020upa} shower, while ``simulation'' uses the default Simple parton shower\cite{Bierlich:2022pfr}. We chose this variation since it ensures sufficiently different distributions where all the differences reside in the physical modeling of soft radiation. We chose {\tt VINCIA} as simulation because we empirically find that the signal and backgrounds are more different, and we want to capture the scenario where the simulations may be too optimistic in distinguishing signal from backgrounds, as the converse is less of an issue.

\section{Bayesian mixture model for di-Higgs searches}\label{sec:bayesian_models}

In this section we provide a brief description of Bayesian mixture models, including inference techniques and metrics for parameter estimation and model comparison. We then build two probabilistic models for the simulated benchmark introduced in Section~\ref{sec:diHiggs}, which are distinguished by whether they infer the number of secondary vertices per jet in an event ($N_{\mathrm{SV}}$) and the distribution of kinematic variables $\mathcal{O}$ or if they assume these probabilities are fixed to the MC estimates. 

Mixture models are ubiquitous in particle physics~\cite{Cranmer:2014lly} in the context of frequentist and/or Bayesian analyses, with broader advances in Bayesian computation becoming more popular in LHC applications in recent years~\cite{Dillon:2019cqt,Fowlie:2019ydo,Dillon:2020quc,Fowlie:2021zyf,DelDebbio:2021whr,Dillon:2021aeo,Brivio:2022hrb,Yallup:2022rjd,Fowlie:2024dgj,Candido:2024hjt,Albert:2024zsh,Fowlie:2025kna,Costantini:2025wxp,Alvarez:2021hxu,Alvarez:2021zje,Alvarez:2022kfr,Alvarez:2022qoz,Alvarez:2024doi,Alvarez:2024owq,Alvarez:2025cnz}. We model an event $e$ as being sampled from a particular class $k$ with probability $\pi_{k}$, with event-level features sampled from the class-dependent distribution $p(e|k,\theta_{k})$ where $\theta_{k}$ are any relevant parameters,
\begin{equation}
    p(e) = \sum_{k=1}^{K} p(e,k)=\sum_{k=1}^{K}\pi_{k}p(e|k,\theta_{k})\,,
\end{equation}
with $\sum_{k=1}^{K}\pi_{k}=1$ and $\int p(e|k,\theta_{k})\, de=1$ for all $k=1,\dots,K$. In this work, $k$ will represent a particular hard process, and we assume that each event originates from a hard process neglecting interference effects.

The key in mixture models lies in the specification of $p(e|k,\theta_{k})$. This requires careful understanding of the data to be analyzed and of the available inference techniques. The data has to be adequately represented, in the sense that the choice of features to record must offer some discrimination power between classes and thus information about the underlying hard process. However, it also needs to be chosen in such a way that $p(e|k,\theta_{k})$ is modellable and inferrable. An arbitrary distribution may be too flexible and thus no inference technique may beat the class degeneracy introduced. This is what happens if, for example, we select $D$ features binned in $\{n_{d}\}_{d=1}^{D}$ bins and try to model each class-dependent distribution as a $\sum_{d=1}^{D}n_{d}-1$ dimensional multinomial distribution. Additional structure needs to be introduced, which produces models that are meaningful and inferrable.

In this work, we follow Refs.~\cite{Alvarez:2024doi,Alvarez:2024owq} and take advantage of the hierarchical structure of the data to separate the features in two sub-sets, one related to the $b$-tagging score and the other to additional kinematic information of the event. As detailed in Section~\ref{sec:diHiggs}, we consider the four leading jets in $p_{T}$ and record their $b$-tagging scores which in this case are the number of secondary vertices per jet $\{N^{i}_{\mathrm{SV}}\}_{i=1}^{4}$. We assume each $N_{\mathrm{SV}}$ follows a flavour- (and crucially, not class-) dependent distribution, which we model as an arbitrary multinomial distribution. We also record a list of kinematic features $\mathcal{O}$ that provide further information about the event, as detailed in Eq.~\ref{eq:kine_features}. To avoid modeling the complex multidimensional distribution $p(\mathcal{O}|k)$, we project it to a one-dimensional variable $x$. That is, we model
\begin{equation}
    p(x|k)=\int \delta(x-f(\mathcal{O}))p(\mathcal{O}|k)\,d\mathcal{O}.\,
\end{equation}
Since $p(x|k)$ is an arbitrary one-dimensional distribution, we bin the observable and consider a multinomial distribution. The key point is that this choice of structure ensures that the different multinomial distributions are learnable, since we reduce a complex distribution to learning a reduced number of class- and flavour-dependent multinomials. To chose the projection and define $x$, we consider the output of a classifier trained on the ``simulation'' dataset. The classifier score approximates a monotonic function of the likelihood ratio in the limit of perfect training. We assume that when considering the same observable in the ``data'' dataset, it will retain a large sensitivity even if it will be no longer optimal due to the differences between the ``simulated'' and ``true'' distributions. Importantly, we assume the classifier to be fixed.

\begin{figure}[t]
    \centering
    \includegraphics[width=0.7\linewidth]{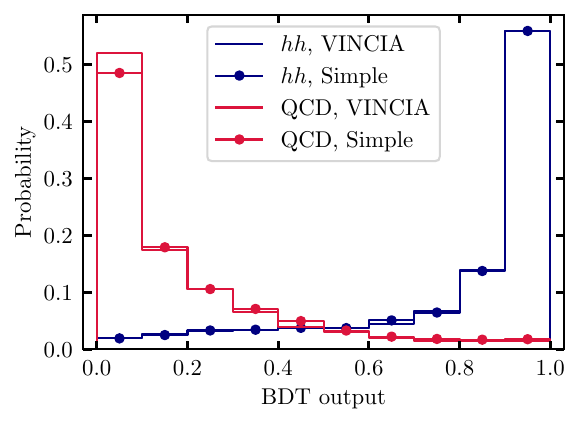}
    \caption{BDT output variable $x$ distribution for the ``simulation'' and ``data'' datasets which correspond, respectively, to Simple and \texttt{VINCIA} histograms. Each entry is an event. The Gradient Boosting Classifier is trained using 10-fold cross-validation. For the ``simulation'' (``data'') samples, the value of $x$ is the average of the remaining 9 (all 10) boosted decision tree (BDT) models. This ensures that no $x$ is obtained by a classifier which was trained on said event.}
    \label{fig:x_sim}
\end{figure}

As a classifier, we consider a Gradient Boosting Classifier trained using the {\tt XGBoost} library~\cite{Chen_2016}, with default hyperparameters except for the learning rate, fixed to 0.3, and the maximum depth, fixed to 5, and which are selected to avoid overfitting while providing satisfactory classification performance. 
We divide the MC dataset on 10 subsets, and train 10 classifiers. The classifier score in ``simulation'' is obtained by averaging the 9 models which have not been trained on that point, and the classifier score in ``data'' is obtained by averaging all 10 classifier scores. We empirically find that this avoids overfitting and provides a feature that remains sensitive to signal and background differences in ``data''.
The resulting BDT output variable $x$ is shown in Fig.~\ref{fig:x_sim} for signal and background in both datasets.

We emphasize that the classifier serves as a projection from a high-dimensional feature space onto a one-dimensional statistic to facilitate Bayesian inference.
Although this projection is designed using MC simulations, its distribution is inferred from data. 
To avoid fitting spurious details of the simulations, it is crucial to employ a classifier of modest complexity—sufficient to capture the relevant structure without fitting artifacts. This approach contrasts with some other studies \cite{Amacker:2020bmn,Chiang:2024pho,Mastandrea:2024irf} that deploy sophisticated neural networks to learn from simulated data.

\begin{figure}[ht]
    \centering
    \includegraphics[width=0.7\linewidth]{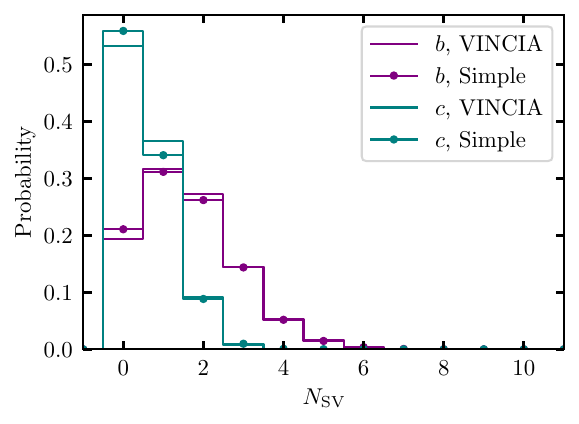}
    \caption{$N_{\rm SV}$ distributions for the ``simulation'' (Simple) and ``data'' ({\tt VINCIA}) datasets, where each entry is a jet.}
    \label{fig:nsv_sim}
\end{figure}

\begin{figure}[ht]
    \centering
    \includegraphics[width=0.35\linewidth]{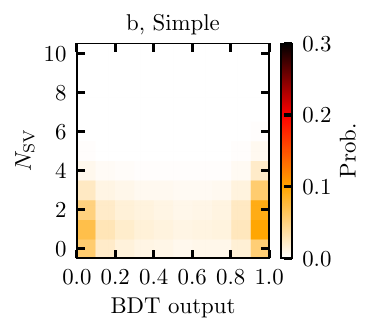}
    \includegraphics[width=0.35\linewidth]{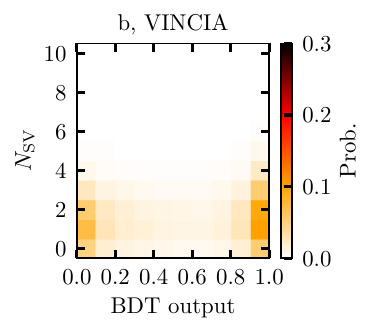} \\
    \includegraphics[width=0.35\linewidth]{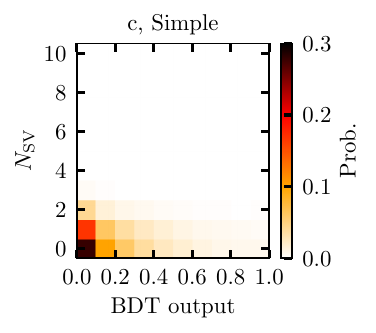}
    \includegraphics[width=0.35\linewidth]{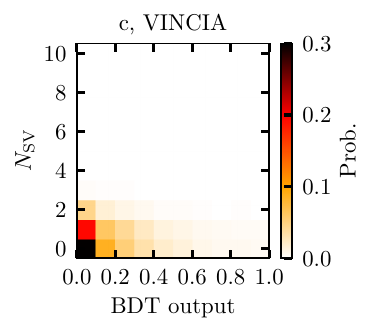}
    \caption{Distribution of event-level BDT output and jet-level $N_{\rm SV}$. Top (bottom) row plots correspond to true b (c) jets, whereas left (right) column plots correspond to Simple (\texttt{VINCIA}) shower model. Each entry is a jet with its corresponding event-level BDT output score.}
    \label{fig:data_correlations}
\end{figure}

In addition to the classifier variable, our model considers the four $N_{\rm SV}$ values corresponding to the hard jets in each event, as defined in Section~\ref{sec:diHiggs}.
The distribution of $N_{\rm SV}$ is given for each flavour and dataset in Fig.~\ref{fig:nsv_sim}, where each entry corresponds to one jet.
As expected, the number of secondary vertices shows some dependence with the kinematics of the event, introducing correlations between the event-level variable $x$ and the jet-level $N_{\rm SV}$, as shown in Fig.~\ref{fig:data_correlations}.
To consider the kinematic effects on the $N_{\mathrm{SV}}$ (and related inter-jet correlations), we condition the $N_{\mathrm{SV}}$ distribution on whether $x$ falls in one of three regions: background-like ($x\in[0,0.2)$), intermediate ($x\in[0.2,0.8)$) and signal-like ($x\in[0.8,1.0]$). We can capture this by introducing a random variable $\mathcal{R}$ such that $\mathcal{R}(x\in[0,0.2))=0,\mathcal{R}(x\in[0.2,0.8))=1,\mathcal{R}(x\in[0.8,1.0])=2$. We empirically found that this splitting is a good compromise between adequately capturing the correlations and having enough statistics per conditional for inference to be possible given the dataset size considered. This is part of the model definition and as thus can be optimized through Bayesian model selection tools.  The model can be written explicitly as,
\begin{equation}
\begin{split}
    p(x,N^1_{\mathrm{SV}},N^2_{\mathrm{SV}},N^3_{\mathrm{SV}},N^4_{\mathrm{SV}}) &=  \sum_{k=hh,4b,2b2c,4c}\pi_{k} p(x,N^1_{\mathrm{SV}},N^2_{\mathrm{SV}},N^3_{\mathrm{SV}},N^4_{\mathrm{SV}}|k)\\
    &= \sum_{k=hh,4b,2b2c,4c}\pi_{k} p(x|k)\sum_{j_i=c,b}p(j_1,j_2,j_3,j_4|k)\prod_{l=1}^{4}p(N^l_{\mathrm{SV}}|j_l,\mathcal{R}(x))\,,
\end{split}
\label{eq:model_1}
\end{equation}
where $\pi_{k}$ are the class fractions, which we can relate to cross sections, $p(x|k)$ are the per-class $x$ kinematic probability distributions, $p(j_1,j_2,j_3,j_4|k)$ is the ``true jet-species'' probability distribution, which we assume known and given by the partonic processes we are considering, and $p(N^i_{\mathrm{SV}}|j_i,\mathcal{R}(x))$ is the $N_{\mathrm{SV}}$ per-jet probability distribution which depends only on the true jet species (not the event class) and the kinematics of the event condensed in $x$. The sum in Eq.~\ref{eq:model_1} is restricted to jet combinations $bbbb$, $bbcc$ and $cccc$, being this an important assumption based on the approximation that the detected jets always correspond to those from the hard process. In Eq.~\ref{eq:model_1} and in the following, we disregard the jet charge since the flavour tagger is agnostic to it and all considered processes have net charge zero at the parton level. Thus, the two jet species are $c$ and $b$ and the final states are called $hh,4b, 2b2c$ and $4c$. For the hard processes we consider, we have
\begin{equation}
    \begin{split}
        p(j_1,j_2,j_3,j_4|hh) &= \prod_{i=1}^{4}\delta_{j_i,b}\,,\\ 
        p(j_1,j_2,j_3,j_4|4b) &= \prod_{i=1}^{4}\delta_{j_i,b}\,,\\ 
        p(j_1,j_2,j_3,j_4|2b2c) &= \frac{1}{6}(\delta_{j_1,b}\delta_{j_2,b}\delta_{j_3,c}\delta_{j_4,c}+\text{ all permutations})\,,\\
        p(j_1,j_2,j_3,j_4|4c) &= \prod_{i=1}^{4}\delta_{j_i,c}\,,\\ 
    \end{split}
    \label{eq:flavour-probs}
\end{equation}
where we ensure at selection that these probabilities are correct. We leave an exploration of more realistic cases where QCD radiation can provide one or more of the four leading jets to future work. 

Another simplifying approximation is that the kinematic distributions for the QCD backgrounds are identical irrespective of the true jet species,
\begin{equation}
    p(x|4b)=p(x|2b2c)=p(x|4c)\equiv p(x|\text{QCD})\,.
\end{equation}
 
\begin{figure}[t]
    \centering
    \includegraphics[width=0.75\linewidth]{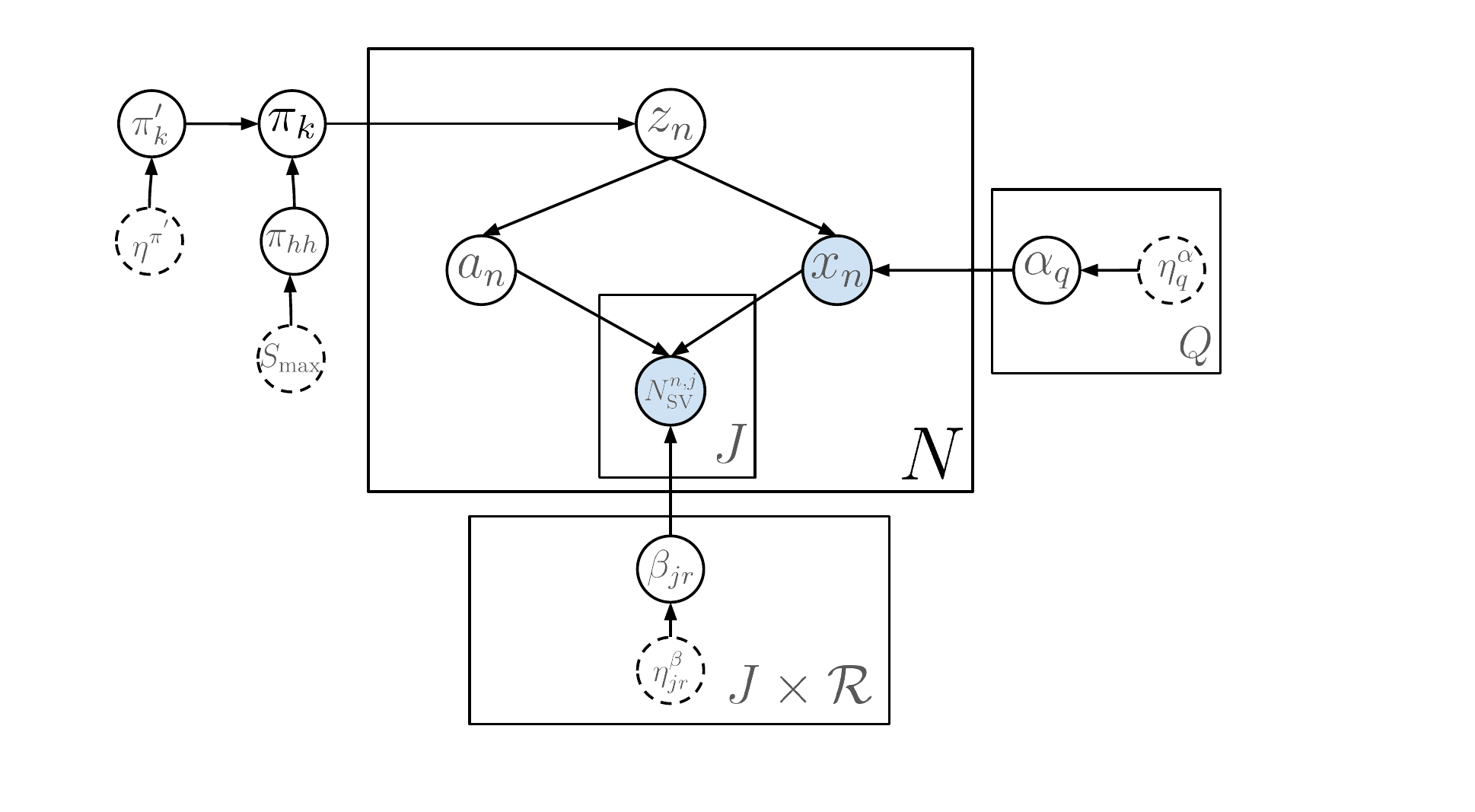}
    \caption{ Graphic model representing the probabilistic model introduced in the main text. White (blue) circles with solid lines represents latent (observed) random variables and circles with dashed lines represent hyperparameters of the model. Box denotes repeated sampling with the number of samples given by the integer at the bottom right corner of the box.}
    \label{fig:graphic_model}
\end{figure}

We bin $x$ in 10 equally sized bins in $[0,1]$ and model the $p(x|q)$ pdf (with $q=hh,\text{QCD}$) as a multinomial where $x\in\text{bin}\,d$ has probability $\alpha_{qd}$ and $\sum_{d=1}^{10}\alpha_{qd}=1$ for all $q$. For $N_{\mathrm{SV}}$ we also consider a multinomial distribution, although here the variable is naturally discrete. If no correlations are considered, we only have one unique set of probabilities per jet-type, $\beta_{jm}$ (where we use $N_{\mathrm{SV}}=m$ for short-hand), while if we consider three $x$ regions we have three sets of conditional probabilities per jet-type $\beta_{jrm}$, such that $\sum_{m=1}^{11}\beta_{jrm}=1$  for each jet-type $j$ and each $x$ region $\mathcal{R}=r$. Inserting these multinomials into Eq.~\ref{eq:model_1} we have that for event $n$,
\begin{equation}
\begin{split}
    &p(x_{n}\in \text{bin }d \text{ and } \mathcal{R} = r,N^{n,1}_{\mathrm{SV}}=m_{1},N^{n,2}_{\mathrm{SV}}=m_{2},N^{n,3}_{\mathrm{SV}}=m_{3},N^{n,4}_{\mathrm{SV}}=m_{4}) = \\
    &\sum_{z_{n}}p(z_{n})p(x_{n}|q(z_{n}))\sum_{a_{n}}p(a_{n}|z_{n})\prod_{i=1}^{4}p(N^{n,i}_{\mathrm{SV}}|x_{n},a_{n})\\
    &=\sum_{k=hh,4b,2b2c,4c}\pi_{k} \alpha_{q(k)d}\sum_{j_i=c,b}p(j_1,j_2,j_3,j_4|k)\prod_{i=1}^{4}\beta_{j_{i}rm_{i}}\,,\\
\end{split}
\label{eq:model_2}
\end{equation}
with the corresponding graphic model shown in Fig.~\ref{fig:graphic_model}, where we have introduced two latent variables, the class-assignment $z_{n}$ can take values $z=hh,4b,2b2c,4c$ with probabilities $p(z=k)=\pi_k$ and the flavour assignments $a_{n}$ which can take values in $\{bbcc,bccb,cbcb,$ $ccbb,bcbc,cbbc\}$ with class-dependent probabilities given by Eq.~\ref{eq:flavour-probs}; and a deterministic variable $q(z)$ such that $q(hh)=hh$ and $q(4b)=q(2b2c)=q(4c)=\text{QCD}$.

Since the number of hard-processes is $K=4$, the number of independent $x$ distributions is $Q=2$, the number of $x$ bins is $D=10$, the number of $x$ regions is $R=3$, the number of jet-types is $J=2$ and $N_{\mathrm{SV}}\in [0,10]$ can take $M=11$ values, we have in total $K-1+Q(D-1)+JR(M-1)=81$ parameters. If we disregard the effect of $x$ in $N_{\mathrm{SV}}$, we have $K-1+Q(D-1)+J(M-1)=41$ parameters. These numbers are much smaller than the naive $(D-1)(M-1)^4 = 9\times10^4$ parameters one would need to characterize $p(x,N^1_{\mathrm{SV}},N^2_{\mathrm{SV}},N^3_{\mathrm{SV}},N^4_{\mathrm{SV}})$, and rely on the appropriateness of $p(j_1,j_2,j_3,j_4|k)$. 

To perform inference, we use Dirichlet priors on each of the feature multinomials, $p(\{\alpha_{qd}\}_{d=1}^{D}|\eta^{\alpha}_{q})=\mathrm{Dir}(\eta^{\alpha}_{q})$ and $p(\{\beta_{jrm}\}_{m=1}^{M}|\eta^{\beta}_{jr})=\mathrm{Dir}(\eta^{\beta}_{jr})$.
The values of the $\eta$ hyperparameters for $\alpha$ and $\beta$ are chosen so that the average distribution matches the expected values under the MC dataset and the overall variance $\Omega^{\alpha,\beta} = \sum \eta^{\alpha,\beta}$ is chosen to reflect a desired spread of the distribution. That is,
\begin{equation}
    \begin{split}
        \eta^{\alpha}_{qd} &= \Omega^\alpha \times \alpha^{\mathrm{MC}}_{qd}\,,\\
        \eta^{\beta}_{jrm} &= \Omega^\beta \times \beta^{\mathrm{MC}}_{jrm}\,,\\
    \end{split}
\end{equation}
where in this work we fix $\Omega^\alpha = \Omega^\beta = 500$ to yield priors that regularize the inference while still allowing for the likelihood term to dominate even in low statistic regions.
The prior for class fractions is given by a two-step sampling process.
First, the signal fraction is sampled from a uniform distribution,
\begin{equation}
    p(\pi_{hh}|S_{\mathrm{max}}) = \mathrm{Unif}\left(0,\frac{S_{\rm max}}{N}\right)\,,
\end{equation}
where $N$ is the total number of events in the sample, and we choose the support interval to allow the range of signal count values from $0$ (absence of signal) to the maximum expected signal events $S_{\rm max}$.
Throughout this work, we fix the hyperparameter $S_{\rm max}=700$.
Then, we introduce the relative background fractions $\pi'$, which are sampled from a Dirichlet distribution, 
\begin{equation}
    p(\{\pi'_{k}\}|\eta^{\pi'})=\mathrm{Dir}(\eta^{\pi'})\,.
\end{equation}
Finally, rescaling enforces the proper normalization, with the background fractions defined as
\begin{equation}
    \pi_k = (1-\pi_{hh})\times\pi'_k\,, \; k\neq hh\,. 
\end{equation}
Similarly to the priors for the feature multinomials, we choose the background relative fraction hyperparameters $\eta^{\pi'}_k=\Omega^{\pi'}\times \gamma^{\mathrm{MC}}_k$ with an overall variance of $\Omega^{\pi'}=70$ and where $\gamma^{\mathrm{MC}}_{k}$ is the relative fraction of background $k$ among all backgrounds in MC simulations.

Having defined priors for all relevant parameters, we characterize the posterior distribution via samples obtained using a Hamiltonian Monte Carlo (HMC) sampler~\cite{betancourt2018conceptualintroductionhamiltonianmonte} implemented in the \texttt{Stan}~\cite{stan} statistical software package. HMC provides a fast, efficient and scalable sampler that is able to take advantage of the derivatives of the posterior with respect to the parameters of interest to explore the relatively high-dimensional parameter space. We have ensured that the sampled posterior chains pass the standard diagnostics of convergence and autocorrelation. 

To evaluate the quality of the model given valid posterior samples, we first inspect visually the posterior distributions. A more qualitative metric is obtained by performing a posterior predictive check\cite{gelman2020bayesian}, where a new dataset is sampled using the parameters from each posterior sample,
\begin{equation}
    \begin{split}
        \pi^{m}, \alpha^{m}_{q},\beta^{m}_{jr} & \sim p(\pi, \alpha_{q},\beta_{jr}|\mathcal{D}^{\mathrm{data}})\,, \\
        \mathcal{D}^{m} &\sim p(\mathcal{D}|\pi^{m}, \alpha^{m}_{q},\beta^{m}_{jr})\,,
    \end{split}
\end{equation}
where $\mathcal{D}^{\mathrm{data}}$ is the original dataset that is used for posterior inference, containing all events $\mathcal{D}=\{e_n\}_{n=1}^{N}$, and $\{\mathcal{D}^{m}\}_{m=1}^{M}$ are M pseudo-datasets of size $N$. Then, a summary statistic $t^{m}=t(\mathcal{D}^{m})$ is obtained from each dataset and we compute the probability distribution of the test statistic $p(t|\mathcal{D}^{\mathrm{data}})$ which can be used in several ways to validate the model. In this case, we focus on the number of sampled signal events $S^{m}$ and study whether the distribution of signal events is consistent with the true value. To express consistency, we compute the $68/95/97\%$ credible intervals defined as the minimum interval  $C_{100\mathrm{f}\,\%}(\mathrm{model},\mathcal{D}^{\rm data})$ that contains $68/95/97\%$ of the samples,
\begin{equation}
    \begin{split}
    C_{100\mathrm{f}\,\%}(\mathrm{model},\mathcal{D}^{\rm data}) &= [S_{\min},S_{\max}]\,, \\
    &= \arg \min_{ [S_{\min},S_{\max}]} (S_{\max}-S_{\min})\quad \text{with} \sum_{S=S_{\min}}^{S_{\max}}p(S|\mathrm{model},\mathcal{D}^{\mathrm{data}})=\mathrm{f} \,,
    \end{split}
\end{equation}
where we have explicitly stated the model dependence implicit in the choice of likelihood and derived posterior in the previous equations. A reasonable model should yield credible intervals that contain the true value\footnote{Additionally, credible intervals yield equivalent results to confidence intervals for models that are accurate enough for the finite dataset statistics.}. 

We work with datasets composed of approximately $14000$ background events and varying $S$. Although we do not weight events by their cross section, the number of background events is meant to reflect the ballpark of what is measurable at the HL-LHC~\cite{atlas4b}. We simulated enough background events that survive the very low acceptance of MC simulation. A more realistic study should increase the simulation size and incorporate the correct cross sections.

\begin{figure}[!ht]
    \centering
    \includegraphics[width=0.49\linewidth]{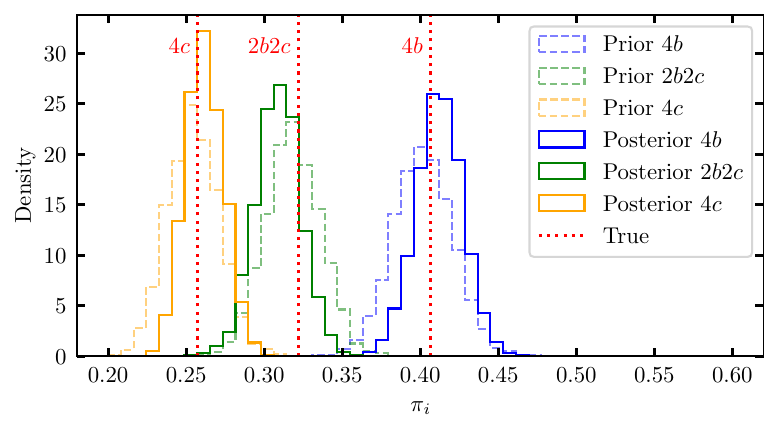}
    \includegraphics[width=0.49\linewidth]{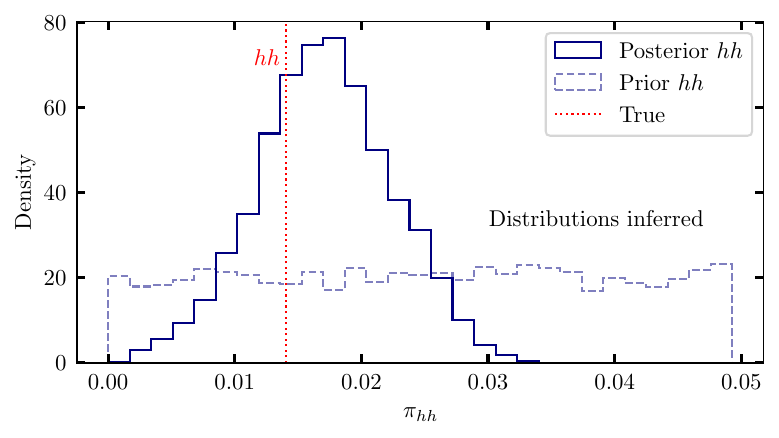}
    \\
    \includegraphics[width=0.49\linewidth]{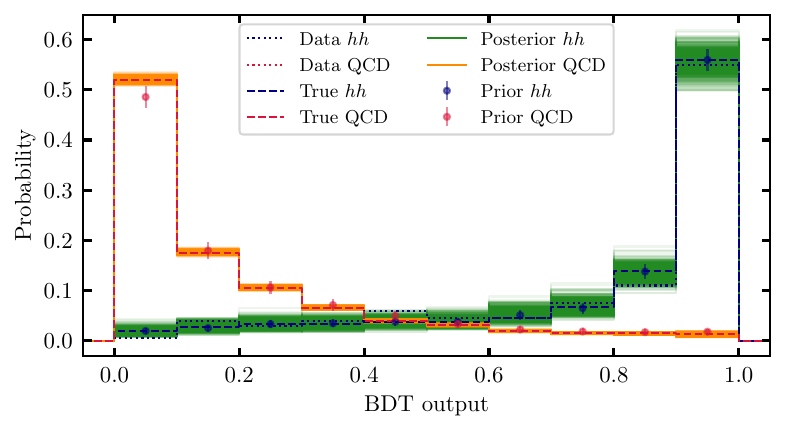}
    \includegraphics[width=0.49\linewidth]{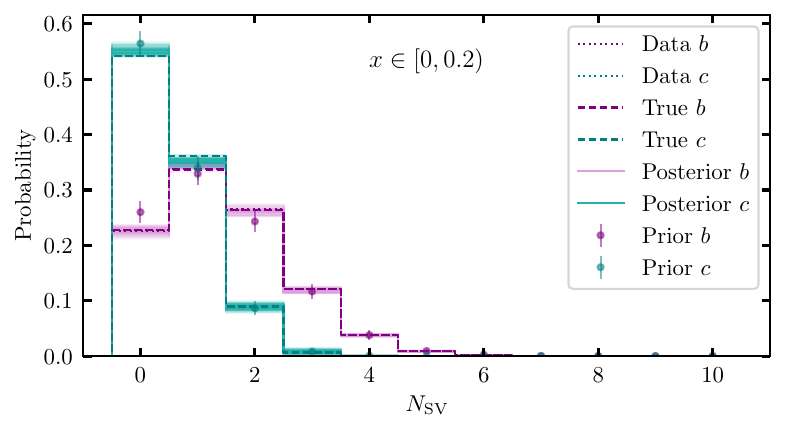}
    \\
    \includegraphics[width=0.49\linewidth]{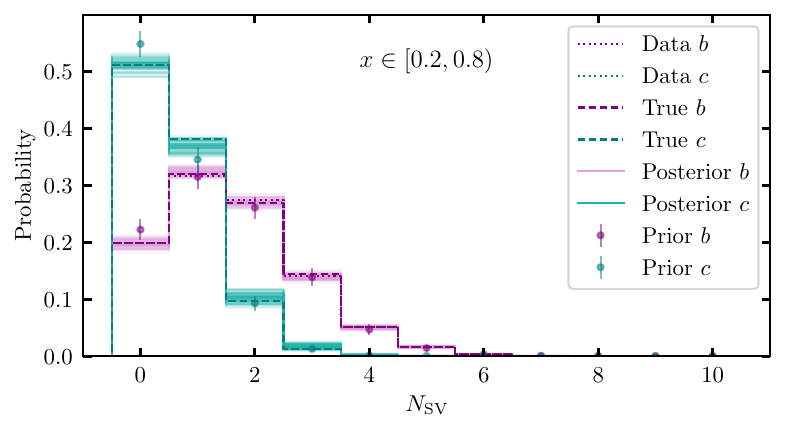}
    \includegraphics[width=0.49\linewidth]{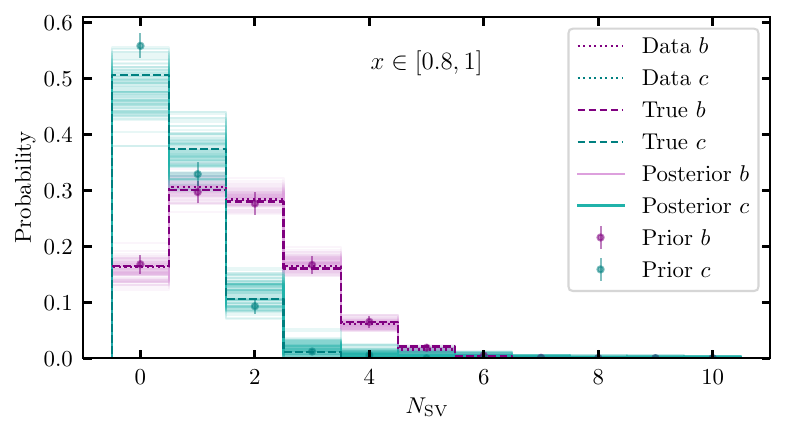}
    \caption{Class fractions and kinematic distributions inferred for $S_{\rm true}=200$. Top row shows samples from the prior and posterior distributions, and true fraction values, for backgrounds (left plot) and signal (right plot). Middle and bottom rows show posterior distributions of the BDT output variable $x$ and $N_{\rm SV}$ for the three kinematic bins. The prior is shown in terms of the mean values and variances of the chosen Dirichlet distributions. Here ``data'' represents the training data, while ``true'' is drawn using all available events to reduce fluctuations.}
    \label{fig:inferred_everything_S200}
\end{figure}

As an example, in Fig.~\ref{fig:inferred_everything_S200} we show the inferred posterior distributions for the class fraction, the BDT output and the number of secondary vertices of each jet as detailed in Section~\ref{sec:diHiggs}. We observe how the posterior distribution correctly shifts closer to the truth, correcting the imperfect prior, and being more compatible with the measured events. We highlight again that we only compare to the truth and measured events per class, but no class information is available during posterior estimation.

We also show the inference results for $S_{\rm true}=100, \,30$ and $0$ in Figs.~\ref{fig:resultsS100},~\ref{fig:resultsS30}~and~\ref{fig:resultsS0} of the Appendix, respectively. 
In presence of signal, we also see how the posterior distribution successfully captures the true distributions.
In the case of no signal, the algorithm correctly learns the true background BDT output, the $N_{\rm SV}$ distributions and background class fractions from  data with only background events. The signal fraction accumulates near $0$, and the learned posterior distribution for the signal BDT output are prior-driven, with small modifications due to the spurious non-zero signal fraction causing the BDT output distribution to be slightly updated using background events. Since true and prior distributions show little mismatch in di-Higgs events, the posterior distribution therefore matches the true distribution within uncertainties, even in the absence of signal events and the influence of background events with non-zero signal probability. 
Since the learned posterior distribution is consistent with the non-zero signal case, and the sculpting of the signal distribution is small, we conclude that the combination of model and priors provide enough inductive bias to be robust against predicting spurious signal events.
More generally, we observe that in all cases the posterior uncertainties reflect the available data, and thus increase in regions where data is limited. 
In particular, for low enough signal fraction, the extreme signal-like region $x\sim 1$ is scarcely populated and thus the relative error increases with respect to low $x$ regions. 
The intermediate $x$ region, although not hugely populated neither by background or signal, is still well-estimated since there are enough background events.

\section{Resilience to Monte Carlo biases}\label{sec:model_comparison}

In this section, we show how Bayesian mixture models are able to learn the appropriate underlying distributions once the proper correlations between features are incorporated. To do so, we compare against a simplified benchmark where $x$ and $N_{\mathrm{SV}}$ distributions are fixed and given by the \texttt{Pythia} Simple parton shower and only the mixing fractions can be inferred.

\begin{figure}[ht!]
    \centering
    \includegraphics[width=0.49\linewidth]{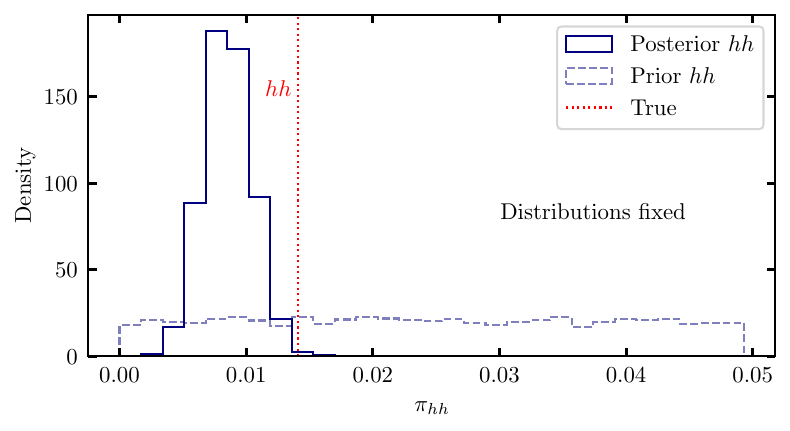}
    \includegraphics[width=0.49\linewidth]{figs/pi1_S200.pdf}
    \\
    \includegraphics[width=0.49\linewidth]{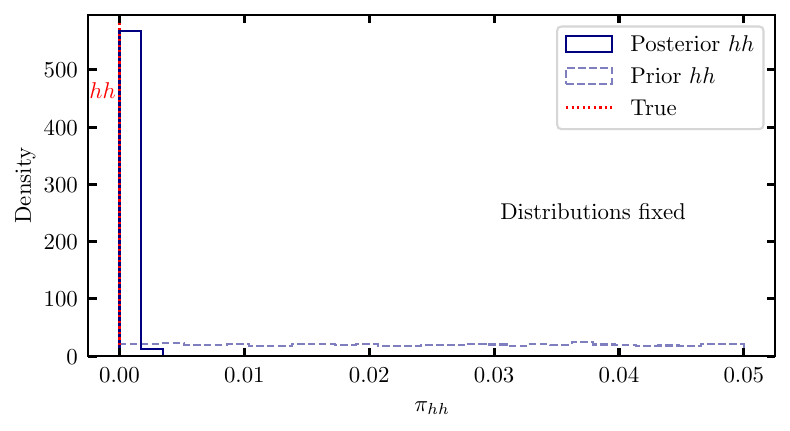}
    \includegraphics[width=0.49\linewidth]{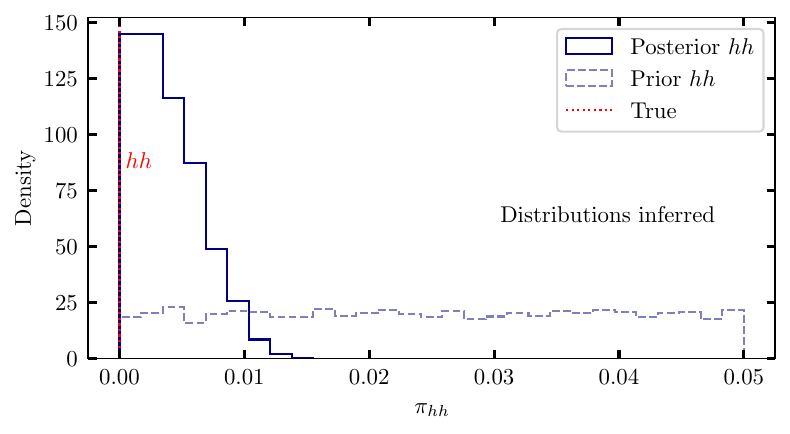}
    \caption{Signal fraction distribution for $S_{\rm true}=200$ and $0$. Inference only over class fractions (left column) and over fractions and kinematic distributions (right column).}
    \label{fig:class_fractions_full_vs_only_fractions}
\end{figure}

The resulting marginal posterior over the signal class fraction is shown in Fig.~\ref{fig:class_fractions_full_vs_only_fractions}. We observe how the model where the kinematic distributions are fixed performs much worse than when inferring the complete set of parameters, with the class fractions being biased for the case of $S\neq 0$. There is an additional effect where the uncertainties are reduced since there is no ``statistical power'' invested in learning the kinematic distribution. This artificially low uncertainty, combined with the bias in the MAP estimator of the class fractions, considerably hinders the performance. The model also fails to estimate the true relative background fractions in this case, as can be seen in Fig.~\ref{fig:fracs_onlyfracs} in the Appendix. The full model where all kinematic distributions are inferred performs better than the model where the kinematic and tagger distributions are fixed to their prior values, and the larger uncertainties reflect the relative prior ignorance on the kinematic distributions and the amount of data used to constrain them.

\section{Framework sensitivity studies}\label{sec:sensitivity}

Having validated the Bayesian framework by verifying that the posterior corrects imperfect prior knowledge of the kinematic distributions by observing the data, we explore in this section how such a posterior-based study can provide useful estimates on the number of signal events and how it compares to other conventional metrics. 

To do so, we perform a posterior predictive check, as detailed in Section~\ref{sec:bayesian_models}, where the test statistic is the number of signal events and for which we provide a credible interval.  In Bayesian statistics, a credible interval is a range containing the most probable values of a parameter, such that the posterior probability that the parameter lies within this range equals a specified probability (e.g. $95\%$). Thus, by finding credible intervals we check whether the model is appropriate in the sense that we test whether the true data is or not unlikely under the posterior probability distribution. 

\begin{figure}[ht]
    \centering
    \includegraphics[width=0.49\linewidth]{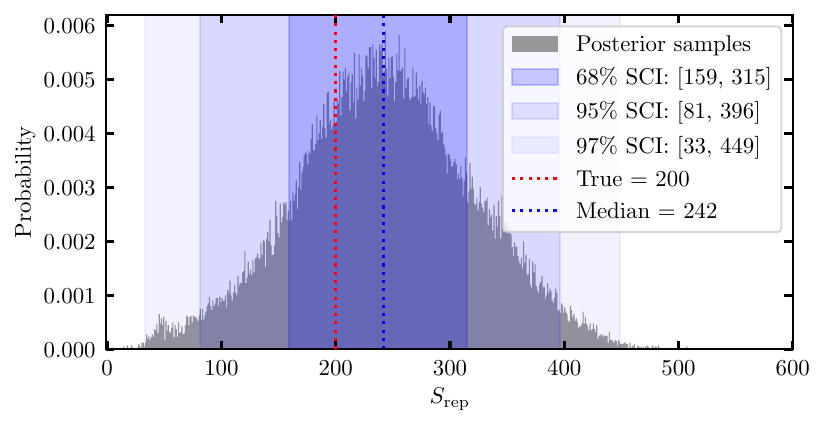}
    \includegraphics[width=0.49\linewidth]{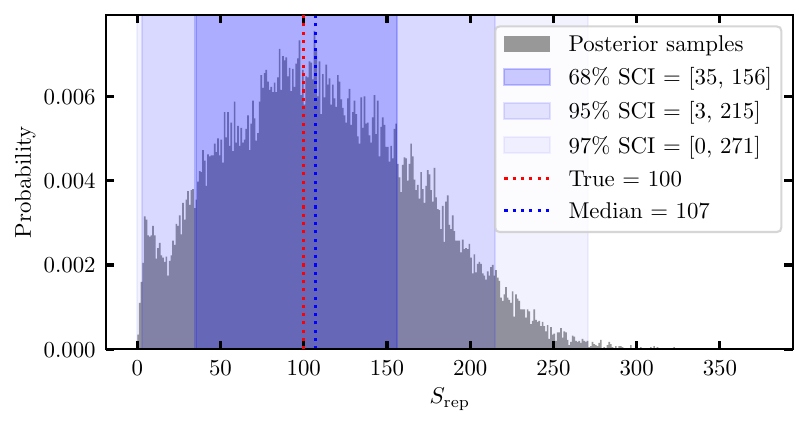}
    \\
    \includegraphics[width=0.49\linewidth]{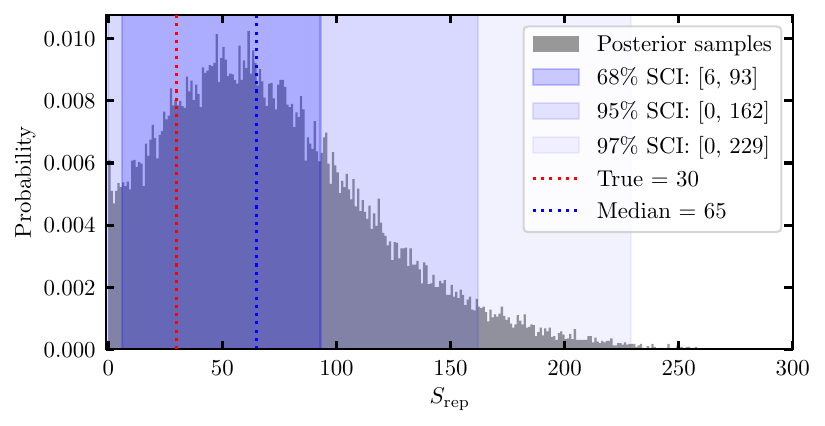}
    \includegraphics[width=0.49\linewidth]{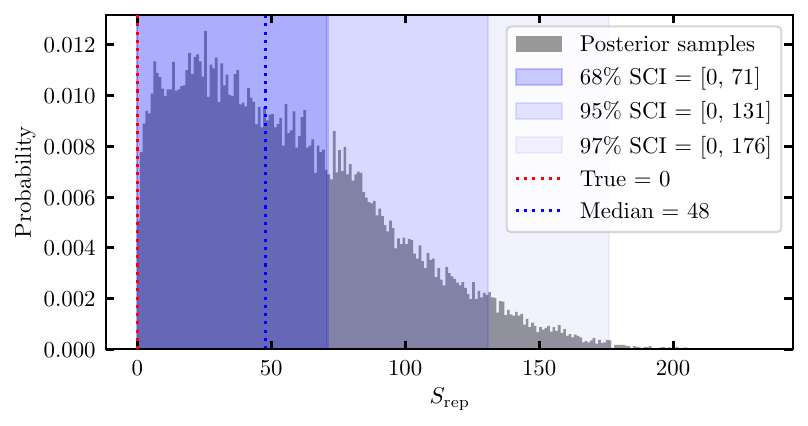}
    \caption{Distribution of number of signal events in data samples from posterior, for $S_{\rm true}=200,100, 30$ and $0$. Median and smallest credible intervals for probabilities of $68\,\%$, $95\,\%$ and $97\,\%$ are shown.}
    \label{fig:credible_intervals}
\end{figure}

The credible intervals for different true signal events are shown in Fig.~\ref{fig:credible_intervals}.  Here we used many samples from the posterior to produce many replica datasets with the same size as the original dataset, and we identify how many signal events there are in each replica.
We observe how the true signal events are never unlikely enough to discard the model. In fact, for all four cases the true number of signal events is contained in the $68\%$ credible interval. For $S=100,30$ and specially $S=0$, boundary effects are important and automatically incorporated in the construction of the credible intervals. The evolution of the distribution between $S=200$ and $S=0$ is continous, allowing to cover multiple regimes with a single prescription to build credible intervals.  As a matter of fact, the resulting distribution on number of signal events in Fig.~\ref{fig:credible_intervals} is a probability {\it mass} distribution whose draws are integer values and contains the zero value as one of the allowed cases.

We observe from inspecting Fig.~\ref{fig:credible_intervals} and the credible intervals, that $S=30$ is close to the minimum injection of signal which becomes compatible with $S=0$ at $68\%$ CL, which can be understood as the minimum sensitivity of the analysis (or, conversely, the upper limit if no signal was present). The impact of boundary effects, and its compatibility to $S=0$, is evidenced in the separation between mean and median of the distribution. 

In addition to the credible intervals, we investigate on the sensitivity that can be achieved by using the results in this work to discriminate the signal-plus-background scenario from the only-background scenario. This is one of the crucial points in searches for di-Higgs events.  Observe that at this level we do not address the question of how much signal there is, or the signal cross-section; we discuss this point in the following paragraphs.  We also compare a proposed method using the current results to customary tools such as the usual $N_S/\sqrt{N_B}$, once it is suitably formalized for its comparison.

Generally speaking, in order to use a given method to discriminate between two possible scenarios or hypotheses, one uses a summary statistic--say $\eta$--whose value determines in which scenario is the system according to the method.  Different thresholds on $\eta$ are referred to as different working points of the method.  In order to assess the discrimination power of the method itself, one can compute the ROC curve (Receiving Operating Characteristic), which is constructed from the distribution of $\eta$ in the two possible hypotheses. Note that here we are referring to a classifier between datasets, not individual events. For independent and identically distributed events, one can construct the former from the latter~\cite{Nachman:2021yvi} but this ceases to be true when uncertainties are taken into account~\cite{Matchev:2020tbw}.

As an example, and to connect to the previously shown credible intervals, we choose $\eta$ to be the number of signal events.  The $\eta$ distribution for different signal-plus-background and only-background scenarios for a fixed posterior is found in Fig.~\ref{fig:credible_intervals}, for a set of replicas obtained from a given posterior distribution. One should observe, however, that a complete hypothesis test should scan over the distribution of posteriors by performing inference on a variety of datasets sampled from the two hypotheses. We do not perform such a test due to the computational cost since we consider that the present version is enough to establish the usefulness of the method. From the $\eta$ distributions we can derive the ROC curve that assesses the discriminative power of this variable by varying the threshold in $\eta$ that determines whether a given dataset is assigned to either the background-only or the signal-plus-background hypotheses.

\begin{figure}[ht]
    \centering
    \includegraphics[width=0.8\linewidth]{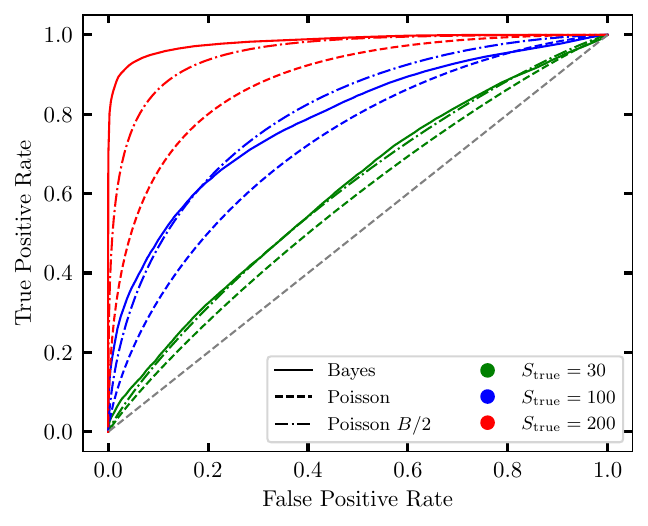}
    \caption{ROC Curves for dataset-level classifiers based on either the sampled number of signal events given the posterior distribution (``Bayes'') or the total number of events sampled from a Poisson distribution (``Poisson''). We show also a cut and count set of curves with reduced background (``B/2'') without affecting the signal, to explore the possible effect of more stringent kinematic cuts.}
    \label{fig:roc}
\end{figure}

For a given threshold in $\eta$ one has a given true positive ratio: given a signal-plus-background scenario, what fraction of all the possible values of $\eta$ lies above the threshold.  Analogously one has a false positive ratio: given an only-background scenario, what fraction of the possible values of $\eta$ lies above the threshold.  We show in Fig.~\ref{fig:roc} the ROC curves for the cases of 30 (solid green), 100 (solid blue) and 200 (solid red) true number of signal events in a background of 14000 events.
The ROC curve can be condensed into the AUC (Area Under Curve), a metric used to assess the discrimination power of a method.
In this case, the AUC is computed using a single number as a summary statistic per replica dataset, as shown in Table~\ref{table:roc}.

As a benchmark to compare against, we consider the simplest summary statistics from a cut and count analysis, the number of events in the sample, namely $\eta = N$.  
In this case, $\eta$ follows a Poisson distribution whose rate depends on the considered hypothesis. The only-background scenario corresponds to $\eta \sim \mbox{Poisson}(N_B)$ whereas for the signal-plus-background scenario corresponds to $\eta \sim \mbox{Poisson}(N_S + N_B)$, where $N_S$ and $N_B$ are the expected central values.
Since the rates are fixed to a particular value, this is also a simplified test which does not account for the complete distribution of possible datasets. A more complete analysis should consider composite null and alternative hypotheses where the rates have an associated uncertainty and thus a prior distribution. However, since we are comparing to a fixed posterior scenario and using the same datasets, we consider this simplified, more optimistic benchmark to be sufficient.
Within this framework one can also compute the ROC curves and their corresponding AUC for the $N_B=14000$ and $N_S=30, 100$ and $200$ cases.  These ROC curves are also plotted in Fig.~\ref{fig:roc} in green, blue and red dashed lines, respectively, and their AUCs are given in the second column of Table~\ref{table:roc}.  
Observe that, for this summary statistic, we recover the usual $N_S/\sqrt{N_B}$ metric, as is expected when comparing the two hypotheses using the likelihood ratio of two Poisson distributions.

\begin{table}[ht]
    \centering
    \begin{tabular}{cccc}
    \hline
    \hline
    $S_{\mathrm{true}}$   & Poisson & Poisson $B/2$ & Bayes \\
    \hline
    30  & 0.57    & 0.60          & 0.61  \\
    100 & 0.72    & 0.80          & 0.79  \\
    200 & 0.88    & 0.95          & 0.98 \\
    \hline
    \hline
    \end{tabular}
    \caption{AUC calculated for the ROC curves shown in Fig.~\ref{fig:roc}.}
    \label{table:roc}
\end{table}

It is worth noting a few points within the above comparison between the posterior-based and Poisson counting summary statistics.  
First, the counting experiment is using the whole sample, without performing any further categorization and/or selection that could increase its performance, as is usually done in realistic analyses.  However, we should keep in mind that we are not using realistic samples, since we are only considering $c\bar cc \bar c$, $b\bar b c\bar c$, and $b\bar b b \bar b$ QCD backgrounds and $b \bar bb \bar b$ signal, and are disregarding important processes such as QCD $jjjj$ or $jjb \bar b$, among others. Thus, the backgrounds are artificially low in this comparison.  The reason for this is that we are interested in a proof of concept of the method, and not a full comparison. Had we taken into account more backgrounds, a more stringent selection could have taken place. This can be done, for example by selecting events where $x>0.5$. In that case, we would get a similar situation as in this work setup. The distribution of signal and background within the $0.5<x<1$ range would be similar to the current distribution (Fig.~\ref{fig:x_sim}), but within a different range. 
A future line of research including more backgrounds and performing a selection that enhances signal over background is feasible but requires considerably more computing resources than the present work.  At the current level we consider that developing this proof of concept, although not a fully fair comparison, is fruitful for showcasing the method and its potential. To provide a more optimistic case for the Poisson counting alternative, we also compute the ROC curves corresponding to the counting Poisson method for the ideal case in which one could cut events in such a way that background events are reduced to half while signal events remain the same, see the dot-dashed ROC curves in Fig.~\ref{fig:roc}, and their corresponding AUCs in the third column in Table~\ref{table:roc}. We observe that, still within this optimistic case, the Bayesian method is competitive to the counting Poisson method.

Another important point to observe in this comparison is that the posterior-based Bayesian method works correcting the MC after seeing the data, whereas in the counting Poisson method its results are dependent on the MC accuracy.  A more detailed comparison should take into account systematic uncertainties. 

Even more, in the above comparison we have used as summary statistics a draw of the posterior of the distribution of number of signal events. However, it would be desirable to use all the information that comes out from the Bayesian framework, which in this case is the whole distribution of number of signal events. That is, the Bayesian method yields a distribution of signal events, which is different for the only-background and the signal-plus-background scenarios. Therefore, a possibly improved method could be as follows. Perform the inference computed in this work many times while varying the MC input for the only-background scenario, and obtain a set of distributions~\footnote{This set can be thought as a distribution of distributions.} of the number of signal events, given a zero true number of signal events. We then use as summary statistics the mean of the KL (Kullback-Leibler) divergence which allows us to compare shapes: i) for the only-background and signal-plus-background scenarios, generate new sets of sampled distributions of the Bayesian method (with repeated posterior inferences for different sampled datasets); ii) use the KL divergence to compare each distribution in these new sets to each distribution in the aforementioned set of only-background scenario distributions; iii) obtain the mean of the KL values over this last set for each distribution in the new sets. Only-background and signal-plus-background scenarios will have different distributions in the summary statistics $\eta=\langle KL \rangle$, and therefore we could compute a new ROC curve for this new method.  Although we may expect a better performance of this new method, we should mention that it relies on varying MC inputs. This method not only entails considerable additional computational resources but also requires a detailed study of MC simulators, which at this level lies beyond the scope of this work.  This approach is a byproduct of the results in this article and should be further investigated, analyzed, and understood.

\section{Outlook}\label{sec:outlook}

We propose a new Bayesian data-driven mixture-model method to study $hh\to 4b$. One novelty of this work is the joint treatment of kinematic and flavour-tagging observables while allowing for dependence between them. The framework implements an interplay between kinematic features, flavour-tagging observables, biased priors, and biased MC that, governed by a model, can infer the correct signal fraction after observing the data.

We deploy this framework on a dataset of 14000 events with QCD backgrounds $4b$, $2b2c$ and $4c$, and a $hh\to 4b$ signal fraction of about $1\%$. We use prior distributions that are biased relative to their true values and let the model learn the true distributions from the data. The Bayesian framework learns in signal region, inferring the mixture parameters by fitting the multidimensional data to the proposed model. We show that the inference procedure can reliably distinguish the no-signal case from the $\sim1\%$ signal case. We find that the model starts to fail to distinguish the no-signal and signal cases for purities of about $0.2\%$ di-Higgs events.

To include kinematic effects of signal and background, we use a simple BDT that maps a variety of kinematic features into a 1D output. We require this BDT to be simple enough to learn the main differences between the classes and not artifacts of the MC simulation. We emphasize this point: controlling the complexity of a multivariate technique helps it learn just what is needed and avoid learning artificial features of a MC in a scenario where we are aware of limitations in reproducing the data. For flavour tagging we use secondary-vertex counting with structured unimodal priors. The interdependence of all variables allows the model to learn the true parameters after observing the data in its multidimensional complexity, despite starting from biased priors.

We summarize the results by computing ROC curves of the proposed framework used as a classifier that distinguishes datasets with di-Higgs events from datasets without di-Higgs events. We find that for 30, 100 and 200 signal events in a sea of 14000 background events, the areas under the ROC curves are 0.61, 0.79 and 0.98, respectively. As a reference, we compare these results against the performance of the customary $N_S/\sqrt{N_B}$ method, yielding improved marks under some simplifying assumptions.

This work is part of a long-term program \cite{Alvarez:2021hxu,Alvarez:2021zje,Alvarez:2022kfr,Alvarez:2022qoz,Alvarez:2024doi,Alvarez:2024owq,Alvarez:2025cnz} to study the implementation of Bayesian techniques in $hh\to 4b$, a promising channel to probe the Higgs self-coupling. Within this program, we intend to show that this channel is currently underestimated because poor knowledge of the overwhelming backgrounds reduces the reach of the currently studied frameworks.
Although we are still at the proof-of-concept level, we can target upcoming objectives toward a concrete proposal of an observable. Among the main goals for future work are the following. One has to add backgrounds such as $4j$, $jjcc$ and $jjbb$ \cite{CMS:2025ero}, with huge cross-sections, as well as incorporate in the analysis the more general flavour structure of the events where the leading jets do not always originate from the hard process. One also has to include single-Higgs backgrounds, such as $bbh$, $tth$ and $Zh$, which, although they have much smaller cross-sections than the QCD ones, look much more similar to the di-Higgs process. 
Although adding these backgrounds and more realistic statistical modeling (also including the effect of different systematics) will increase the sophistication and associated computing cost of the analysis, we expect that its implementation will not differ qualitatively from the one presented here. 
However, that analysis will allow a meaningful estimation of the sensitivity of our method in a more realistic scenario that matches the number of events expected for the HL-LHC.

It is worth mentioning that as we make progress within this program on Bayesian techniques in $hh\to 4b$, we accomplish the proposed objectives. Some of the goals proposed in previous works \cite{Alvarez:2024owq,Alvarez:2024doi} have been achieved in this work, namely: the incorporation of more realistic complex dependence between the observed variables and b-tagging observable derived from jet features, the inclusion of the QCD four-bottom background, and the simultaneous inference of kinematic and flavour-tagging observables.

Finally, we remark a promising finding of this work. As a summary statistic to classify the signal versus no-signal datasets, we used samples of the inferred number of signal events in the datasets. However, we are not completely using the available information contained in the samples: exploiting the {\it shape} of the posterior in the number of signal events should enhance the discrimination power of the classifier. This objective requires not much more than scaling the computing power to obtain probability distributions over posterior shapes upon varying the MC tunes and the possible measurable datasets; that is, one needs to study in more detail the mapping between biased MCs and posterior shapes. We stress that the latter is obtained after observing the data--which is sampled under a given hypothesis--within the proposed interplay and framework. 
We therefore expect an enhancement from this new method for classifying datasets, which could help to compensate expected difficulties coming from the addition of backgrounds and the accounting of a more realistic multi-jet structure in the model.

In conclusion, the proposed Bayesian mixture-model framework provides a principled, data-driven way to combine kinematics and flavour tagging under imperfect prior knowledge and/or biased simulations, achieving robust inference down to percent-level signal purities. With the planned extensions and HL-LHC-scale studies, this approach can serve as a complementary analysis strategy to traditional searches and may help unlock the full potential of the $hh\to 4b$ channel for Higgs self-coupling measurements.

\section*{Acknowledgments}
We express our gratitude to the public universities and the state research organizations of Argentina for their enduring commitment in the face of ongoing challenges.
MS acknowledges support in part by the DOE grant DE-SC1019775, and the NSF grants OAC-2103889, OAC-2411215, and OAC-2417682.
ST acknowledges the hospitality and support of the Institut ``Jožef Stefan'' and the Institut de Physique Théorique during the research stays undertaken in the course of this work.

\appendix

\section{Additional figures for Bayesian inference}

\begin{figure}[!ht]
    \centering
    \includegraphics[width=0.49\linewidth]{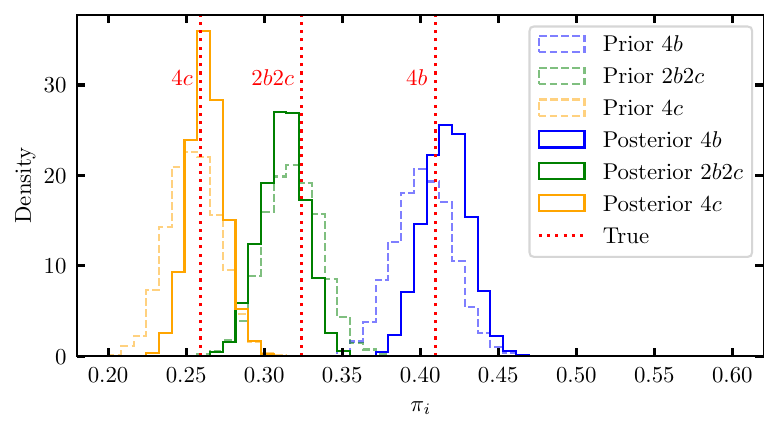}
    \includegraphics[width=0.49\linewidth]{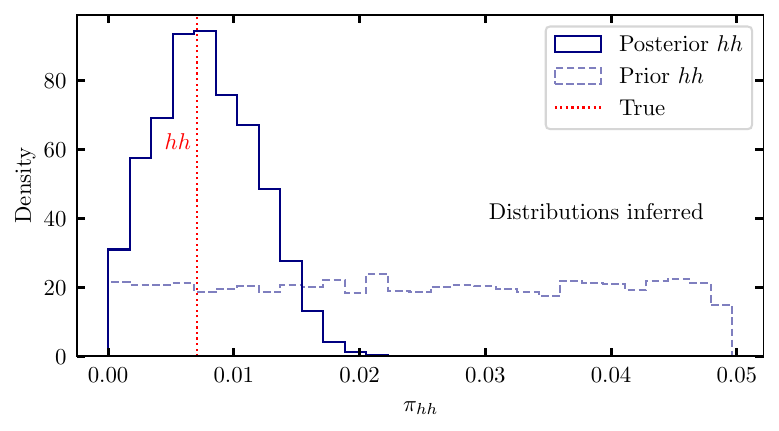}
    \\
    \includegraphics[width=0.49\linewidth]{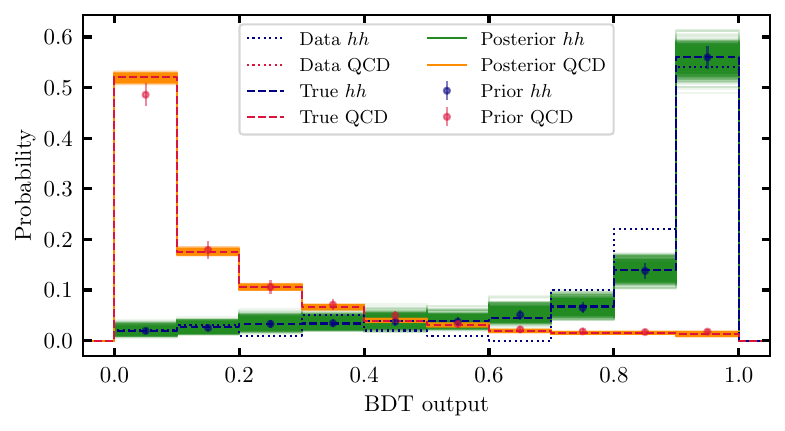}
    \includegraphics[width=0.49\linewidth]{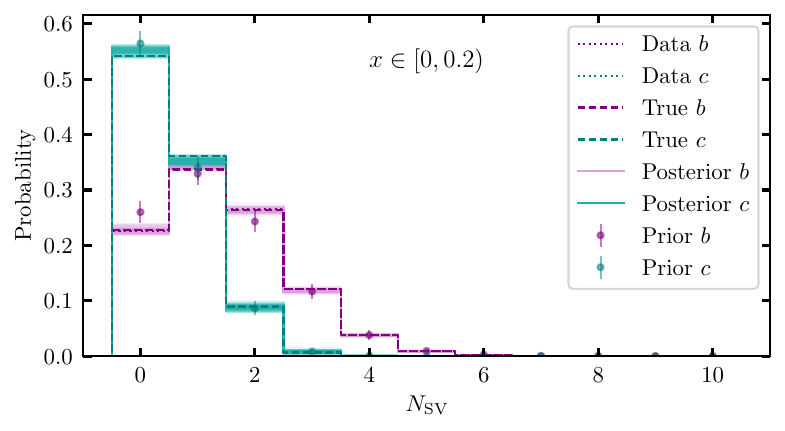} 
    \\
    \includegraphics[width=0.49\linewidth]{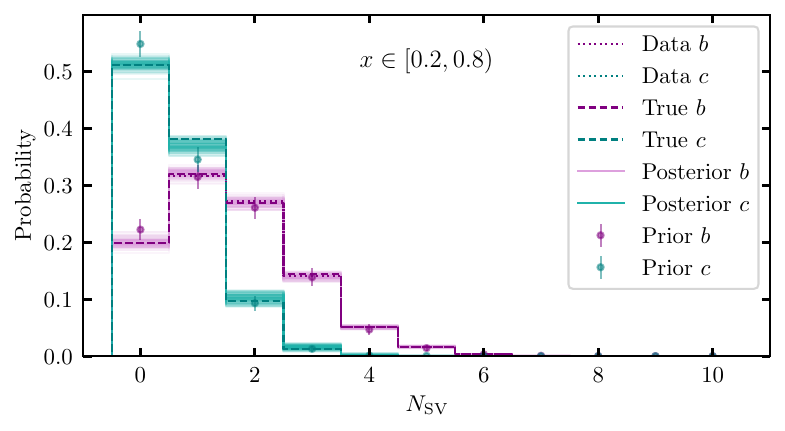}
    \includegraphics[width=0.49\linewidth]{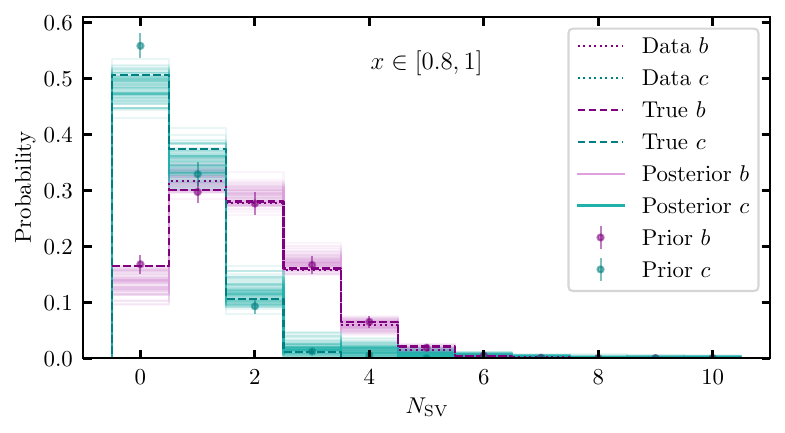}
    \caption{Class fractions and kinematic distributions for $S_{\rm true}=100$}
    \label{fig:resultsS100}
\end{figure}

\begin{figure}[!ht]
    \centering
    \includegraphics[width=0.49\linewidth]{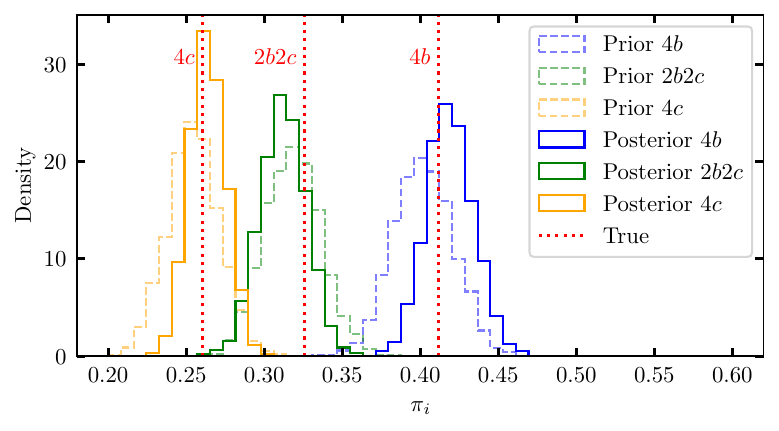}
    \includegraphics[width=0.49\linewidth]{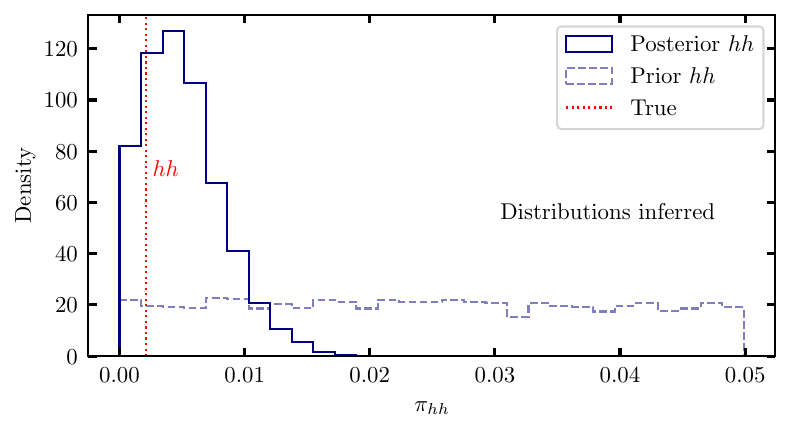}
    \\
    \includegraphics[width=0.49\linewidth]{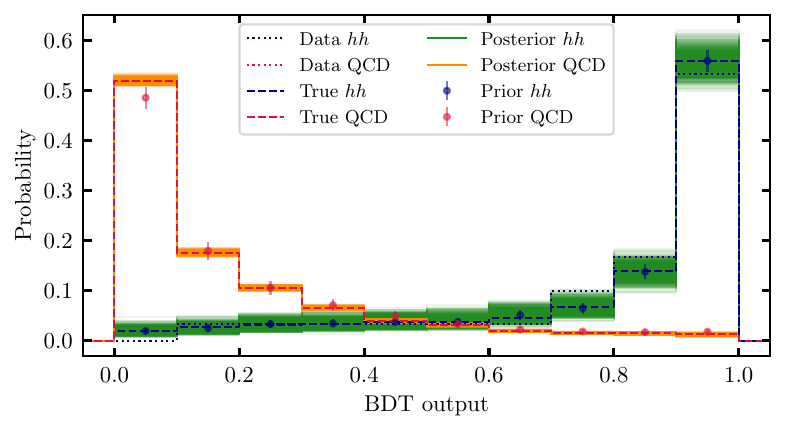}
    \includegraphics[width=0.49\linewidth]{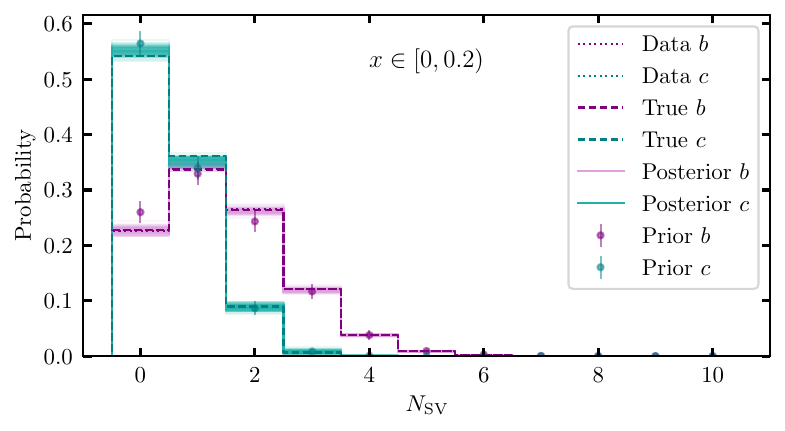} 
    \\
    \includegraphics[width=0.49\linewidth]{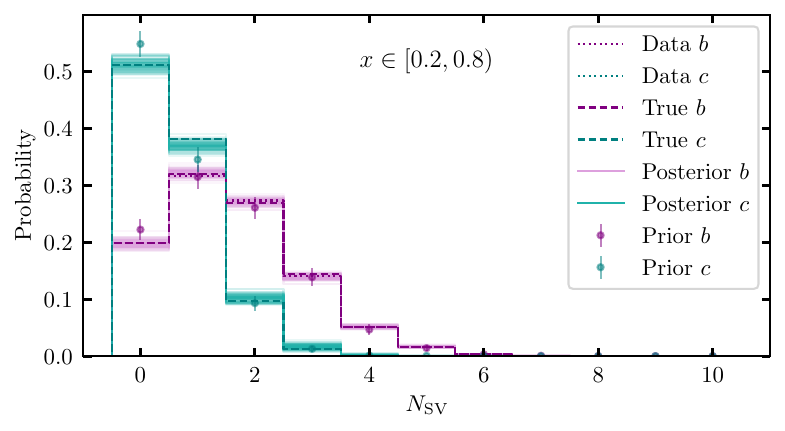}
    \includegraphics[width=0.49\linewidth]{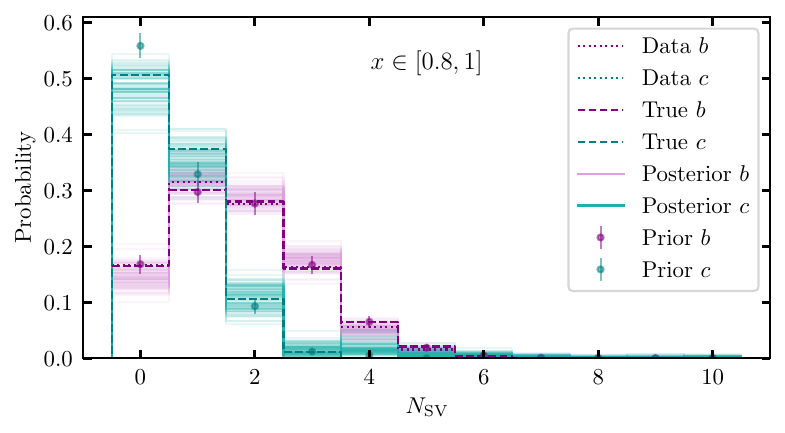}
    \caption{Class fractions and kinematic distributions for $S_{\rm true}=30$}
    \label{fig:resultsS30}
\end{figure}

\begin{figure}[!ht]
    \centering
    \includegraphics[width=0.49\linewidth]{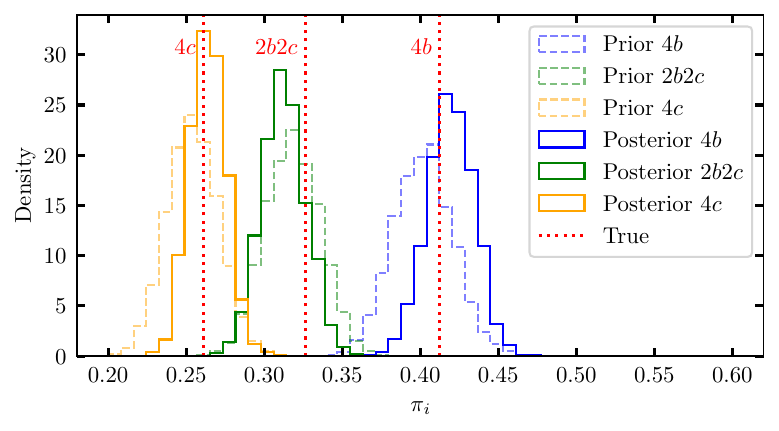}
    \includegraphics[width=0.49\linewidth]{figs/pi1_S0.pdf}
    \\
    \includegraphics[width=0.49\linewidth]{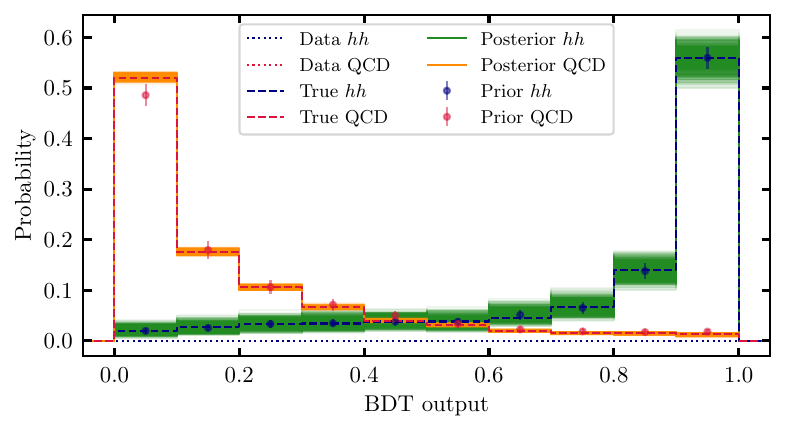}
    \includegraphics[width=0.49\linewidth]{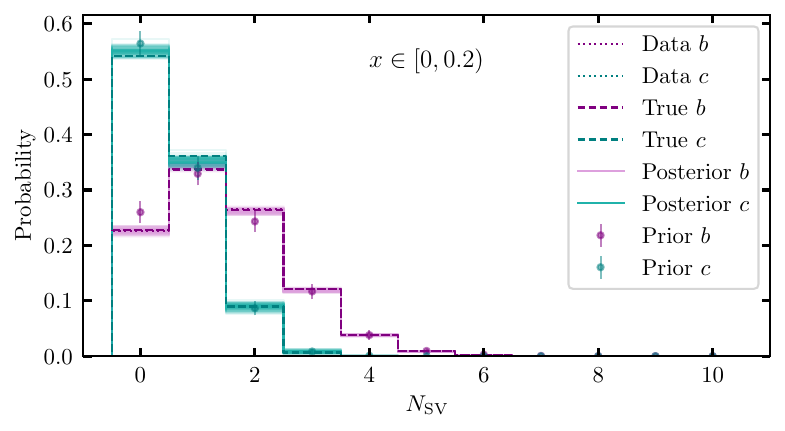}
    \\
    \includegraphics[width=0.49\linewidth]{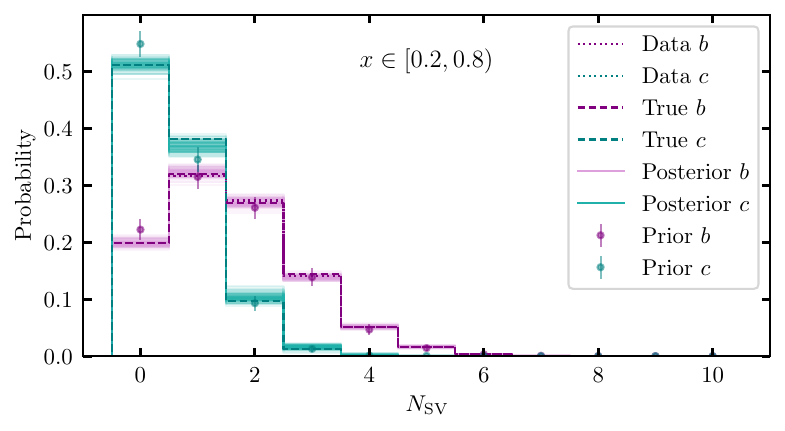}
    \includegraphics[width=0.49\linewidth]{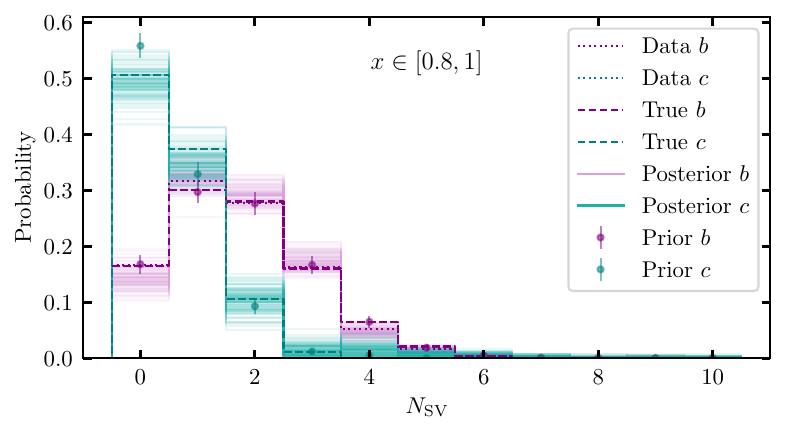}
    \caption{Class fractions and kinematic distributions for $S_{\rm true}=0$}
    \label{fig:resultsS0}
\end{figure}

\begin{figure}
    \centering
    \includegraphics[width=0.49\linewidth]{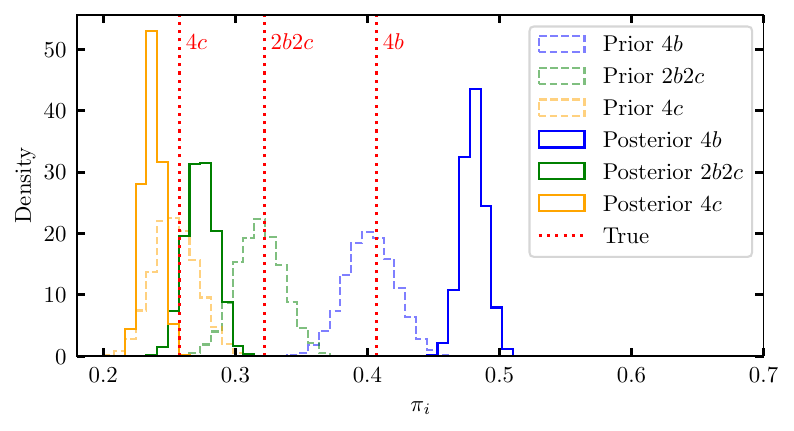} 
    \includegraphics[width=0.49\linewidth]{figs/onlyFrac_pi1_S200.pdf} 
    \\
    \includegraphics[width=0.49\linewidth]{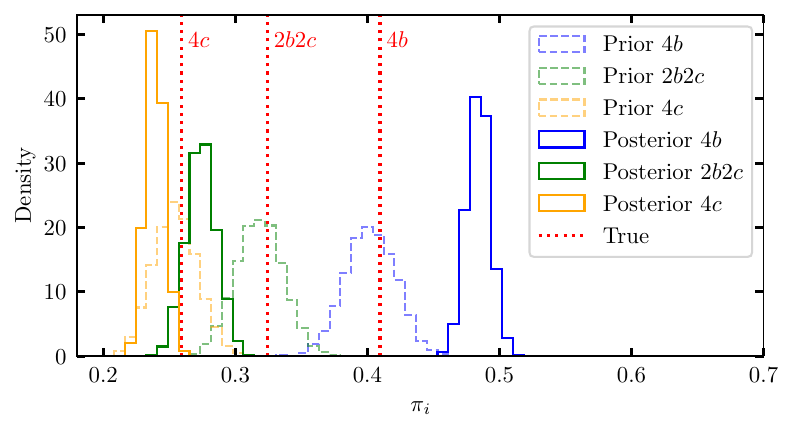}
    \includegraphics[width=0.49\linewidth]{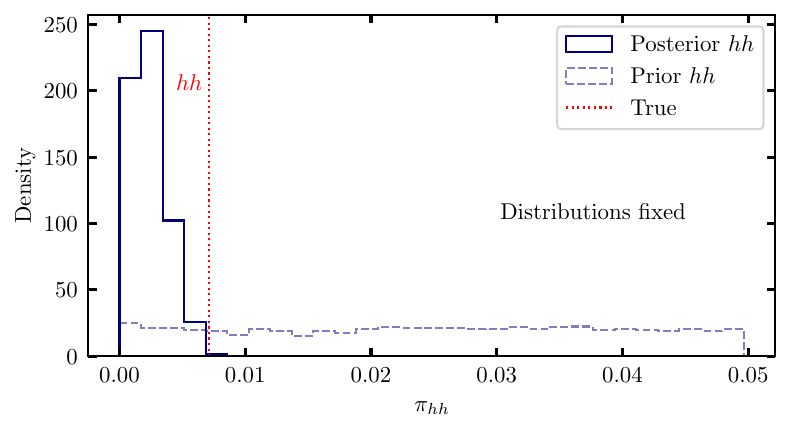}
    \\
    \includegraphics[width=0.49\linewidth]{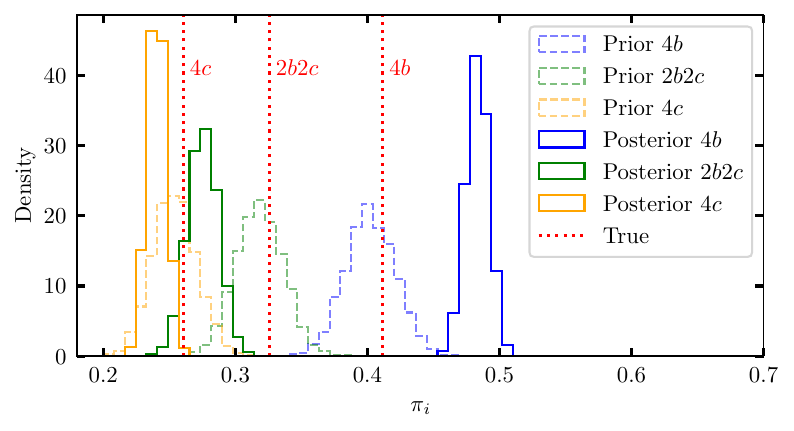}
    \includegraphics[width=0.49\linewidth]{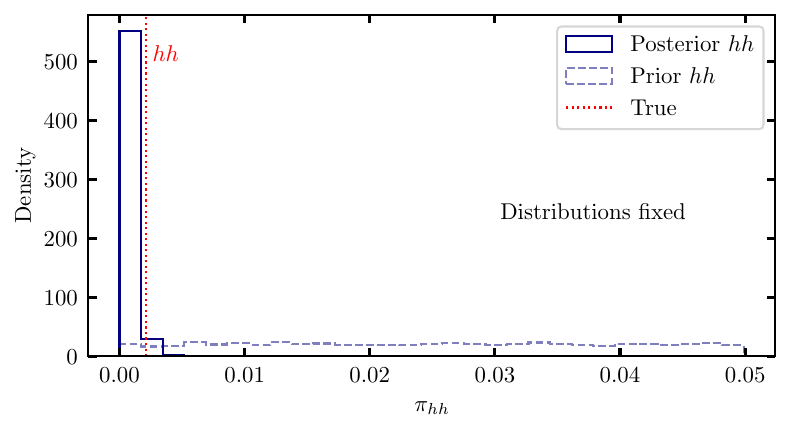}
    \\
    \includegraphics[width=0.49\linewidth]{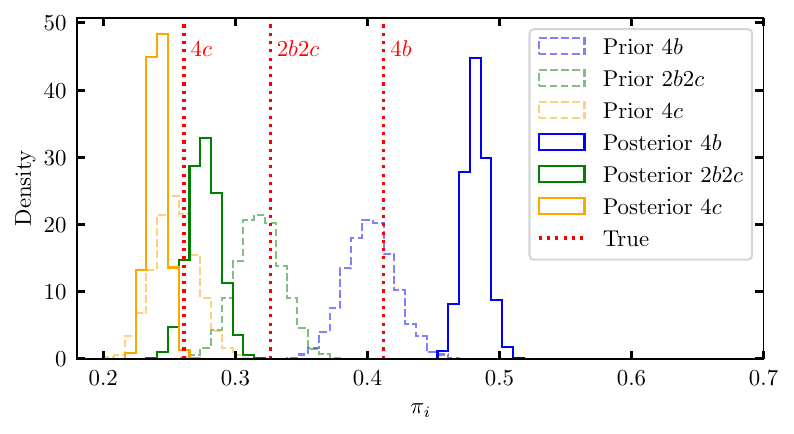}
    \includegraphics[width=0.49\linewidth]{figs/onlyFrac_pi1_S0.pdf}   
    \caption{Distribution of posterior samples of class fractions for backgrounds (left) and signal (right), when inference is done with kinematic distributions fixed. Rows correspond from top to bottom to $S_{\rm true}=200,\,100,\,30$ and $0$.}
    \label{fig:fracs_onlyfracs}
\end{figure}

\begin{figure}[!ht]
    \centering
    \includegraphics[width=0.49\linewidth]{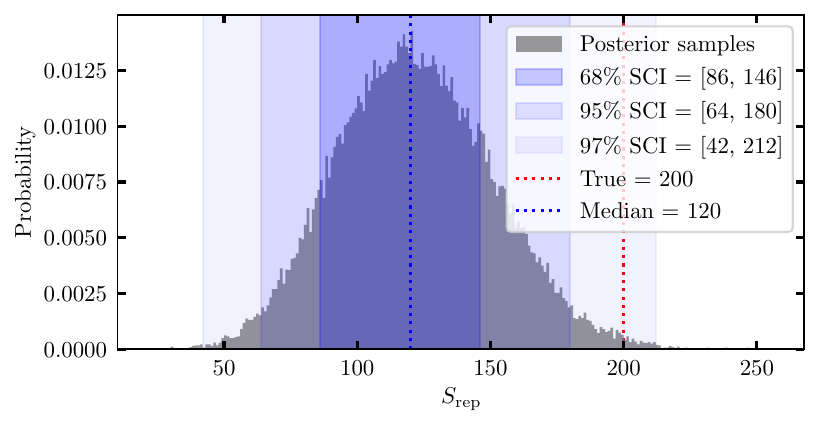}
    \includegraphics[width=0.49\linewidth]{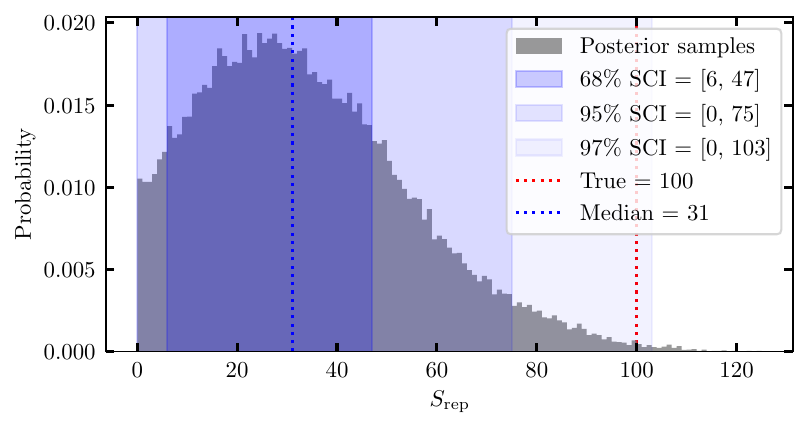}
    \\
    \includegraphics[width=0.49\linewidth]{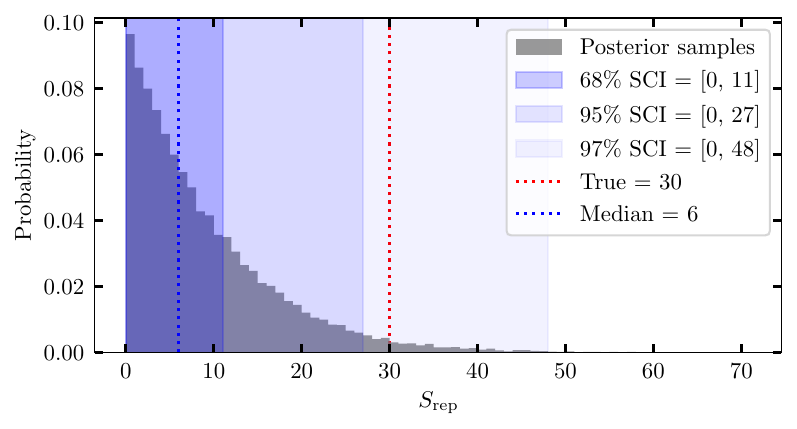}
    \includegraphics[width=0.49\linewidth]{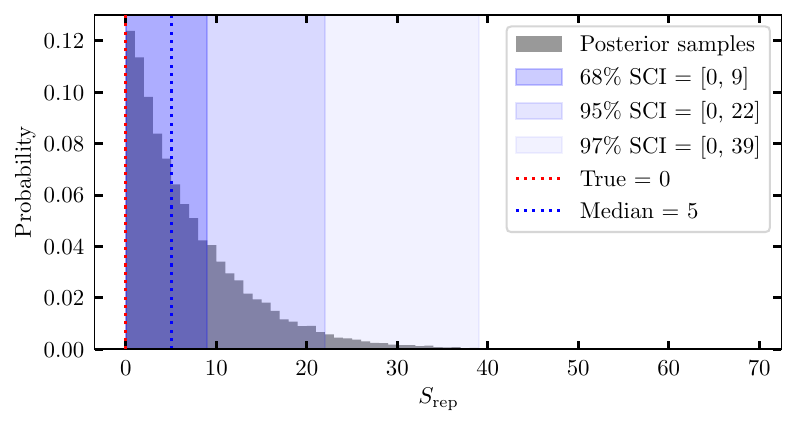}
    \caption{Distribution of number of signal events in data samples from posterior, for $S_{\rm true}=200, 100, 30$ and $0$ when only class fractions are inferred. Median and smallest credible intervals for probabilities of $68\,\%$, $95\,\%$ and $97\,\%$ are shown.}
    \label{fig:ppc_onlyfrac}
\end{figure}

We show in Figs.~\ref{fig:resultsS100},~\ref{fig:resultsS30} and~\ref{fig:resultsS0} the inference results, analogous to Fig.~\ref{fig:inferred_everything_S200}, for the cases corresponding to $S_{\rm true}=100,\,30$ and $0$, respectively.
We also show plots corresponding to the case where we keep kinematic distributions fixed to prior mean values and we only infer class fractions, which are shown in Fig.~\ref{fig:fracs_onlyfracs} for $S_{\rm true}=200,\,100,\,30$ and $0$.
Finally, in Fig.~\ref{fig:ppc_onlyfrac} we show the distributions of the number of signal events in posterior samples for this case.

\clearpage
\bibliography{biblio}

\end{document}